\documentclass[twocolumn]{aastex631}

\usepackage{diagbox}
\usepackage{amsmath}
\usepackage{graphicx}
\usepackage{subfigure}
\usepackage{longtable}
\usepackage[figuresright]{rotating}
\usepackage{orcidlink}

\begin{document}

\def\liu#1{{\bf #1}}

\shorttitle{Massive Quiescent Disk Galaxies at $0.5\ {\leq}\ z\ {\leq}\ 1$ in CANDELS}
\title{Massive Quiescent Disk Galaxies at $0.5\ {\leq}\ z\ {\leq}\ 1$ in CANDELS : Color Gradients and Likely Origin}
\email{E-mail: fsliu@nao.cas.cn}

\author{Qifan Cui\orcidlink{0009-0001-5320-1450}}
\affil{National Astronomical Observatories, Chinese Academy of Sciences, 20A Datun Road, Chaoyang District, Beijing 100101, China}
\affil{Key Laboratory of Space Astronomy and Technology, National Astronomical Observatories, Chinese Academy of Sciences, 20A Datun Road, Chaoyang District, Beijing 100101, China}

\author{Pinsong Zhao}
\affil{National Astronomical Observatories, Chinese Academy of Sciences, 20A Datun Road, Chaoyang District, Beijing 100101, China}
\affil{Key Laboratory of Optical Astronomy,
National Astronomical Observatories, Chinese Academy of Sciences, 20A Datun Road, Chaoyang District, Beijing 100101, China}

\author{F. S. Liu $^{\color{blue} \dagger}$}
\affil{National Astronomical Observatories, Chinese Academy of Sciences, 20A Datun Road, Chaoyang District, Beijing 100101, China}
\affil{Key Laboratory of Optical Astronomy,
National Astronomical Observatories, Chinese Academy of Sciences, 20A Datun Road, Chaoyang District, Beijing 100101, China}
\affil{School of Astronomy and Space Science, University of Chinese Academy of Sciences, Beijing 100049, China}

\begin{abstract}
A rare population of massive disk-dominated quiescent galaxies has recently drawn much attention, which intrudes the red sequence population without destroying the underlying stellar disks. In this study, we have carefully identified 48 red sequence (RS), disk-dominated galaxies with $M_{\ast} > 10^{10}M_{\odot}$ between redshift 0.5 and 1.0 in all five CANDELS fields. These galaxies are well fitted by a two-component bulge plus disk model, and have the bulge-to-total ratio $B/T<0.4$ in the both F814W and F160W bands. The fitting results indicate that these galaxies generally have extended stellar disks ($\rm \sim 3~kpc$ on average) and tiny bulge components ($\rm \sim 0.5~kpc$ on average). To understand their possible origins, we have also selected two control samples of 156 green valley (GV) and 309 blue cloud (BC) disk-dominated galaxies according to the same selection criteria. We study the $UVI$($U-V$ versus $V-I$) color gradients of these galaxies to infer their specific star formation rate (sSFR) gradients out to the maximum acceptable radii. We show that on average the disks in disk-dominated RS galaxies are fully quenched at all radii, whereas both the BC and GV disks are not fully quenched at any radii. We find that all the BC, GV and RS disk galaxies generally have nearly flat sSFR profiles. We propose a potential formation mechanism, acknowledging that various other mechanisms (e.g., central compaction and AGN feedback) might contribute, where massive quiescent disk-dominated galaxies are predominantly formed via a process of secular disk fading.

%


\end{abstract}

\section{Introduction} \label{sec:intro}

It has been known that galaxies display bimodality in both optical color and morphology in the color-magnitude (stellar mass) and morphology-stellar mass diagrams both locally and at high redshifts \citep[e.g.,][]{Kauffmann2003MNRAS.341...54K,Blanton2003ApJ...594..186B,Baldry2004ApJ...600..681B,Bell2004ApJ...608..752B,Faber2007ApJ...665..265F,Brammer2009ApJ...706L.173B,Ilbert2010sf2a.conf..355I}. Galaxies are broadly divided into two classes: (1) star-forming systems with prominent disks often exhibiting blue color and spiral structure. (2) quiescent systems dominated by a spheroidal morphology-the most extreme being ellipticals-usually having very little ongoing star formation and red color. Understanding the processes by which galaxies transform from the ``blue cloud" of star-forming to the "red sequence" of quiescent systems has been a key topic in both observational and theoretical studies of galaxy evolution over the past several decades. The common scenario is that galaxy quenching processes are accompanied with a structural transition from star-forming disks to quenched spheroids. However, in addition to the dominated population of spheroidal galaxies at the massive end ($M_{\ast} > 10^{10}M_{\odot}$) of the red sequence, a rare population of red disk galaxies also intrude the red sequence. The existence of such massive red (quiescent) disk galaxies challenges the current standard paradigm of galaxy formation \citep[][]{Xu2021SCPMA..6479811X}.

Red disk galaxies have actually been known for several decades \citep[e.g.,][]{vandenBergh1976ApJ...206..883V, Dressler1999ApJS..122...51D, Poggianti1999ApJ...518..576P}. Red optical color may result from attenuation due to dust \citep[][]{Valentijn1990Natur.346..153V,Driver2007MNRAS.379.1022D,Masters2010MNRAS.404..792M}, high metallicity \citep[][]{Mahajan2009MNRAS.400..687M}, low star formation rate \citep[][]{Moran2006ApJ...641L..97M,Masters2010MNRAS.405..783M,Cortese2012A&A...543A.132C,Tojeiro2013MNRAS.432..359T}, or no detectable star formation \citep[][]{Fraser-McKelvie2016MNRAS.462L..11F}. 
It has been shown that inclined star-forming disk galaxies may appear red due to 
dust reddening \citep[e.g.,][]{Masters2010MNRAS.404..792M}. However, some researches suggest that 
numerous red disk galaxies are really quiescent and have very little ongoing star formation \citep[e.g.,][]{Masters2010MNRAS.405..783M}. 
%
It has raised great concern about how quiescent disk galaxies formed and were quenched.

In the local universe, various properties of quenched massive disk (spiral) galaxies have been revealed, and many efforts have recently been invested in understanding of their possible origins. For instance, based on the Galaxy Zoo project, \citet{Masters2010MNRAS.405..783M} found that at high stellar masses ($M_{\ast} > 10^{10}M_{\odot}$), a significant fraction of spirals are truly quiescent disks and the environment effect is not sufficient to quench these massive disks. In consideration of the old stellar populations hosted by red spirals and the intact disk morphology, they proposed that red spirals might be old spirals that have exhausted all of their gas. \citet{GuoRui2020ApJ...897..162G} performed a statistical analysis of the spectroscopic and structural properties of local red spiral galaxies with $M_{\ast} > 10^{10.5}M_{\odot}$ in SDSS DR7, and compared them with ellipticals and blue spiral galaxies of similar mass. 
They showed that massive red spiral galaxies have compact cores, 
large Dn(4000) values, high [Mg/Fe] values, and dark matter halo masses similar to elliptical galaxies, which suggest that the bulges of red spirals formed within a short timescale before redshift $z \sim 1-2$ and were quenched via a fast mode. They also found that about 70\% of the red spiral galaxies show strong bar, ring/shell structures and even merger features, based on the optical morphology. \citet{Hao2019ApJ...883L..36H} carried out a two-dimensional spectroscopic analysis of a subsample of \citet{GuoRui2020ApJ...897..162G} out to 1.5 times effective radii based on the SDSS DR15 for the MaNGA survey. They found that the stellar population properties of their red spirals are more similar to those of red elliptical galaxies than to blue spiral galaxies, and explanation of this is that red spirals form as remnants of gas-rich major mergers occurring above $\rm z\sim1$. \citet{Zhang2019ApJ...884L..52Z} found that nearly all massive quiescent disk galaxies still retain cold atomic hydrogen gas similar to star-forming galaxies, but have significantly reduced molecular gas and dust content and lower star formation efficiency. \citet{Mahajan2020MNRAS.491..398M} showed that the optical colors of red spiral galaxies at $\rm 0.002<z<0.06$ in GAMA are a direct consequence of some environment-related mechanism(s) that has removed dust and gas, leading to a lower star formation rate. \citet{Zhou2021ApJ} investigated the star formation histories of red spiral galaxies with $M_{\ast} > 10^{10.5}M_{\odot}$, and make comparisons with blue spirals and red ellipticals of similar masses, using MaNGA spectra. They suggested that massive red spirals are likely to share some common processes of formation with massive red ellipticals in the sense that both types were formed at $\rm z>2$ through a fast formation process. \citet{Shimakawa2022PASJ...74..612S} showed the proportion of passive spiral galaxies is $\sim2\%$ at $\rm 0.01<z<0.3$ by deep imaging data taken from the HSC-SSP PDR3 over 1072 $\rm deg^2$. They also found that the characteristics of stellar populations in passive spirals were similar to those of typical quiescent galaxies.

At moderate and high redshifts, \citet{Bundy2010ApJ...719.1969B} examined the evolution of passive spirals since $\rm z\sim1-2$ based on the COSMOS survey and discussed extensively the possible origins of red disk galaxies. They showed that passive disk galaxies are a significant component of red sequence galaxies, with about 60\% of galaxies transitioning onto the red sequence experiencing a passive disk phase. There is also evidence for massive galaxies with low bulge-fractions and already ancient stellar populations at $\rm z\sim1-3$ \citep[e.g.,][]{McGrath2008ApJ,Stockton2008,vandW2011ApJ,Bruce2014MNRAS,2014AAS...22314515K}. More recently, \citet{Fudamoto_2022} detected two extremely red spiral galaxies likely in the cosmic noon ($\rm 1<z<3$) as well, from the JWST ERO image data release of SMACS 0723, one of which is more likely to be a passive galaxy having moderate dust reddening. The presence of massive disks of old stars at high redshifts means that at least some quiescent disk galaxies have formed directly through highly dissipative events with significant angular momentum in the early universe.

It is still unclear how massive quiescent disk galaxies formed, and how the star formation was quenched without destroying the underlying stellar disks. Different selection criteria have resulted in mutually exclusive samples of optically red disk galaxies with distinguishable properties. For instance, \citet{Masters2010MNRAS.405..783M} selected disk-dominated spirals, rejecting bulge-dominated disk galaxies, whereas most of red spiral galaxies in \citet{GuoRui2020ApJ...897..162G} have the bulge-to-total ratio $B/T>0.5$, harboring a prominent bulge. On the other hand, massive quiescent disks are likely a collection of several individual populations that are a product of different quenching mechanisms or effects \citep{Mahajan2020MNRAS.491..398M}. There are various physical processes responsible for cessation of star formation in galaxies, such as central compaction \citep{Fang2013ApJ...776...63F,Dekel2014MNRAS.438.1870D,Zolotov2015MNRAS.450.2327Z,Tacchella2016MNRAS.458..242T,Liu2016ApJ...822L..25L,Liu+18,Barro2017ApJ...840...47B,Whitaker2017ApJ...838...19W}, AGN feedback \citep{Croton2006MNRAS.365...11C,Fabian2012ARA&A..50..455F,Harrison2017NatAs...1E.165H}, supernova feedback \citep{Geach2014Natur.516...68G}, major merger \citep{Mihos1996ApJ...464..641M,Hopkins2008ApJS..175..356H}, environmental effects \citep[][]{Kauffmann2004MNRAS.353..713K,Baldry2006MNRAS.373..469B,Peng2012ApJ...757....4P,Geha2012ApJ...757...85G,Guo2017ApJ...841L..22G,Papovich2018ApJ...854...30P}, halo quenching \citep{Dekel2006MNRAS.368....2D}, morphological quenching \citep{Martig2009ApJ...707..250M}, gravitational quenching \citep[][]{Genzel2014ApJ...785...75G}, and strangulation \citep[][]{Larson1980ApJ...237..692L,Peng2015Natur.521..192P}. Different mechanisms are expected to change the radial sSFR profile of a galaxy in a different way during its quenching process. For example, the processes of central compaction and AGN feedback first deplete the gas in the centers of galaxies or blow it out of the centers, 
causing ``the inside-out'' quenching. The environmental effects usually strip the gas content of galaxies first from their outskirts, causing ``the outside-in'' quenching. In the case of quenching by ``strangulation'', star formation can continue with the gas available in the galaxy until it is completely used up, which may cause a uniform decline of the radial sSFR profile from the center to the outskirts.

In this work, we focus exclusively on quiescent disk-dominated galaxies with low 
bulge-to-total ratios at moderate redshifts.
To investigate their possible origins, we have carefully identified 
48 red sequence (RS), 156 green valley (GV) and 309 blue cloud (BC) disk-dominated galaxies 
with $M_{\ast} > 10^{10}M_{\odot}$ between redshift 0.5 and 1.0 in all five CANDELS fields.  
we exploit deep high-resolution HST/WFC3 and Advanced Camera for Surveys (ACS) multi-band imaging data 
to measure their rest-frame $UVI$ color profiles to infer their sSFR profiles. We show that on average 
the RS disks are fully quenched at all radii, whereas both the BC and GV disks 
are not fully quenched at any radii. We find that all the BC, GV and RS disks generally 
have nearly flat sSFR profiles, and that the GV disks have more similar structural properties 
to the RS disks than to the BC disks. We discuss possible meachanism(s) for the formation of 
massive quiescent disks.

The outline of this paper is as follows.
In Section \ref{sec:data and sample}, we describe the data and sample selection.
In Section \ref{sec:identificaiton of disks}, we describe the identification of disk-dominated
galaxies based on the bulge/disk decomposition with {\tt GALFIT}.
In Section \ref{sec:galaxy classification}, we describe the classification for the RS, GV and BC 
disk galaxies. 
In Section \ref{sec:color profiles}, we describe the measurement of color profiles.
We present our main results in Section \ref{sec:results and analysis}
and finish with a discussion and summary in Section \ref{sec:Discussion and Conclusion}.
Throughout the paper, we adopt a cosmology with $\Omega_{M}$=0.3,
$\Omega_{\Lambda}$=0.7 and $\rm H$=70$~km~s$$^{-1}$~Mpc$^{-1}$.
All magnitudes are in the AB system.

\section{Data and Sample} \label{sec:data and sample}

The data and sample selection procedure used in this work are roughly the same as those of \citet{Liu+18}. We here present a brief description and refer readers to \citet{Liu+18} for details. In order to maximize the sample size, we select massive galaxies from all five CANDELS fields \citep{Grogin2011ApJS,Koekemoer2011ApJS,Faber2011}. Multi-band photometry catalogs were built by \citet[for COSMOS]{Nayyeri2017ApJS}, \citet[for EGS]{Stefanon2017ApJS}, \citet[for GOODS-N]{Barro2019ApJS}, \citet[for GOODS-S]{Guo2013ApJS} and \citet[UDS]{Galametz2013ApJS}, respectively. 
Redshifts used in this work are the best redshifts, which are in the order of priority of 
secure spectroscopic, good grism, and photometric redshifts if available.
Spectroscopic redshifts were recently recompiled by N. P. Hathi (2018, private communication) for all five CANDELS fields, which include publicly available \citep[e.g.][and reference therein]{Santini2015ApJ} and unpublic (e.g., UCR DEIMOS Survey) redshifts. Grism redshifts came from the 3D-HST/CANDELS Survey \citep[][]{Morris2015AJ, Momcheva2016ApJS}. Photometric redshifts were estimated using the multiwavelength photometry catalogs and adopting a hierarchical Bayesian approach \citep{Dahlen2013ApJ}. Rest-frame integrated magnitudes from FUV to K were computed using {\tt EAZY} \citep{Brammer2008ApJ}, which fits a set of galaxy SED templates to the multi-wavelength photometry, with the redshifts as inputs. Stellar masses were computed using {\tt FAST} \citep{Kriek2009ApJ} and based on a grid of \citet{Bruzual2003MNRAS} models that assume a \citet{Chabrier2003PASP} initial mass function (IMF), exponentially declining $\tau$-models, solar metallicity, and a \citet{Calzetti2000ApJ} dust law. SFRs were computed from rest-frame UV luminosities at $\lambda$ $\approx$ 2800 Å that are corrected for extinction by applying a foreground-screen Calzetti reddening law \citep{Calzetti2000ApJ,Kennicutt2012ARA&A}. We adopted the median $\rm A_V$ that was calculated median by combining results from four methods \citep[see labeled 2a$\tau$, 12a, 13a$\tau$, and 14a in][]{Santini2015ApJ} if available. Effective radius ($R_e$) along the semi-major axis ($R_{\rm SMA}$), minor-to-major axis ratio ($\rm q$) and position angle ($\rm PA$) were measured from the F160W images using {\tt GALFIT} \citep{vanderWel2012ApJS}. 

\begin{table*}[htbp]
\normalsize
\caption{Sample Selection Criteria and Sample Sizes for CANDELS \label{tab:table_1}}
\begin{tabular}{ccccccc}
\hline
\diagbox{Criterion}{Field} & COSMOS & EGS & GOODSN & GOODSS & UDS & TOTAL\\ 
\hline
Full Catalog & 38671 & 41457 & 35445 & 34930 & 35932 & 186435 \\
$\rm Hmag {\leq} 24.5$ & 11811 & 11292 & 9522 & 8352 & 9758 & 50735  \\
$\rm GALFIT\ Flag(H) = 0$ & 9110 & 9031 & 7770 & 6698 & 8090 & 40699\\
$\rm CLASS\_STAR < 0.9$ & 8861 & 8817 & 7660 & 6577 & 7960 & 39875\\
$\rm 0.5 {\leq} z {\leq} 1$ & 3119 & 2485 & 2544 & 2147 & 2404 & 12699\\
$\rm M_* {\geq} 10^{10}M_{\odot}$ & 669 & 415 & 469 & 353 & 325 & 2231\\
$\rm R_{SMA} {\geq} 0.''18$ & 602 & 383 & 420 & 335 & 294 & 2034\\
$\rm PhotFlag=0$ & 318 & 198 & 403 & 330 & 288 & 1537\\
Good multi-band data & 224 & 189 & 324 & 250 & 206 & 1193\\
\hline
Disk-dominated galaxies & 104 & 91 & 132 & 97 & 89 & 513\\
RS disk  galaxies & 6 & 6 & 16 & 11 & 9 & 48\\ 
GV disk  galaxies & 33 & 28 & 34 & 36 & 25 & 156\\ 
BC disk  galaxies & 65 & 57 & 82 & 50 & 55 & 309\\ 
\hline
\end{tabular}
\end{table*}

The parent sample used in this work is constructed by applying the following cuts to the above data:

\begin{enumerate}
\item Observed F160W($H$) magnitude brighter than 24.5 to ensure high signal-to-noise ratios (S/Ns). 
\item {\tt GALFIT} quality $\rm flag = 0$ (good fit) in F160W \citep[][]{vanderWel2012ApJS} to ensure well-constrained measurements of structural parameters. 
\item {\tt SExtractor} $\rm CLASS\_STAR < 0.9$ to exclude stars. 
\item Redshifts within $0.5 {\leq} z {\leq} 1$ and stellar masses $\rm M_{\ast} {\geq} 10^{10}M_{\odot}$. 
\item $R_{\rm SMA}{\geq}0.18^{\prime\prime}~(\rm 3~drizzled~pixels)$ to minimize the point spread function (PSF) effects on color gradient measurement. 
\item No contamination from blending (PhotFlag=0 and visual inspection), the borders of the mosaic, or bad pixels to interfere with the fitting. Available simultaneous measurements of the surface brightness profiles (SBPs) in $F606W$, $F814W$, $F125W$ and $F160W$. 
\end{enumerate}

Table \ref{tab:table_1} details our selection criteria and the resulting sample sizes after each cut for each field. After the above cuts, we obtain 1193 massive galaxies in total from all five CANDELS fields.

\section{Identification of Disk Galaxies} \label{sec:identificaiton of disks}

\begin{figure*}
\centering
\includegraphics[width=1.0\textwidth]{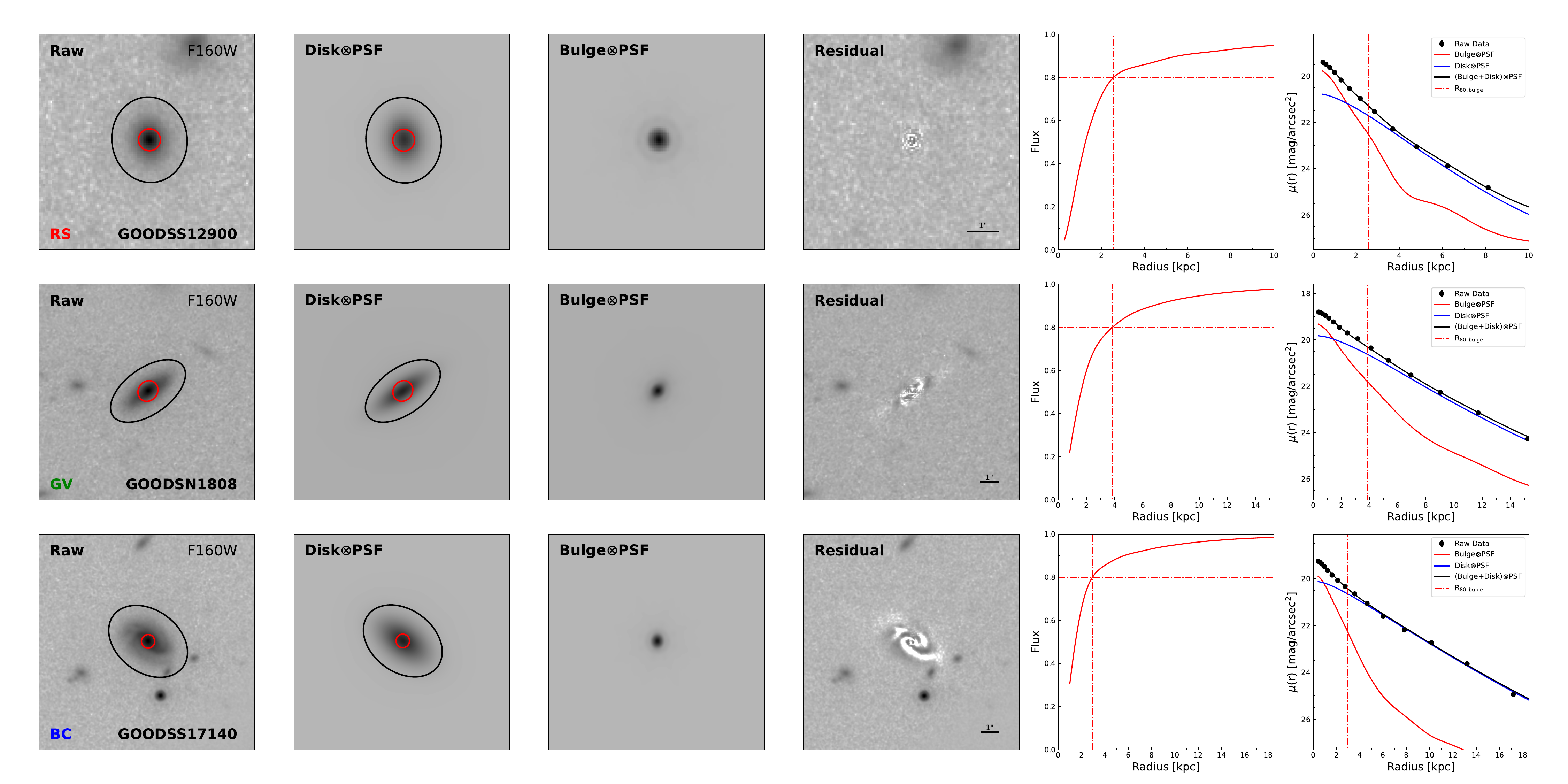}
\caption{\label{fig:Figure_1}
Best {\tt GALFIT} fittings with the two-component SerExp model for three examples of RS, GV and BC 
galaxies (from top to bottom), respectively. 
From left to right, panels in the first column show the raw images in F160W.
Panels in the second and third columns show the disk and bulge components of two-component SerExp model, respectively.
The black circles show the Kron apertures measured by {\tt SExtractor}, and the red circles indicate
the $R_{80,bulge}$. 
Panels in the fouth column show the residual images after subtracting the two-component SerExp model
from raw images. Panels in the last two columns show the growth curves of the bulge components
and the comparisions between the observed and the best fitted model profiles, respectively. 
}
\end{figure*}


Galaxy morphology measurements were obtained using {\tt GALFIT} \citep{Peng2002}. 
We first generated the multi-band thumbnail images from the original CANDELS mosaics for individual targets. 
The thumbnails have a size of 401 $\times$ 401 pixels. A careful sky background subtraction 
was then applied to each thumbnail. For this purpose, we made segmentation maps with 
{\tt GNUASTRO} \citep{Akhlaghi2015ApJS..220....1A} and developed a pixel-by-pixel background estimation code 
in Python. We produced the noise maps by adding the Poisson noise of individual galaxies to the root mean square 
(rms) maps, which include the detector and background noises. 
We provided a PSF to {\tt GALFIT} that was constructed by stacking 
the unsaturated stars with high $S/Ns$ carefully-selected in individal field. 
To mask bad pixels and exclude interference 
from neighbouring sources, we first ran {\tt SExtractor} dual model to generate segmentation maps of each thumbnail. 
The mask maps were then generated by setting the region of target galaxy to be zero. 
A 2.5 times Kron radius measured by {\tt SExtractor} in F160W was used to constrain the size of 
{\tt GALFIT} fitted region. The initial structural parameters prior to {\tt GALFIT} were 
determined with {\tt SExtractor}.

We fit two models for both the F160W and F814W band thumbnail images of each selected galaxy: 
the two-component model of a S\'ersic bulge with a variable $n$ plus an exponential disk (SerExp), 
and the single-component S\'ersic model with a variable index ($n$). 
For the two-component SerExp model, we set the following constraints: the disk $n$ is fixed at 1.0, 
the bulge $n$ ranges from 0.5 to 8.0, and the $R_e$ ranges from 0.5 to 50 pixels. In addition, we restrict 
$R_e$ (disk)/$R_e$ (bulge)$>$1, which indicates that the identified disk component is the outer disk. 
For the single-component S\'ersic model, we use the same constraints as for the bulge of the two-component SerExp model.

After obtaining the best-fitting results with both the two-component SerExp model and 
the single-component S\'ersic model, we selected galaxies with the bulge-to-total ratios $B/T<0.4$ by 
the two-component SerExp model in both the F160W and F814W bands as our disk-dominated galaxies. 
Specially, for the galaxies with $B/T<0.05$ in either of these two bands, we instead used the fitting results 
by the single-component exponential disk model (S\'ersic $n=1$) in the following analysis.       
Finally, we identify a total of 513 massive disk-dominated galaxies in all five CANDELS fields. 
Figure \ref{fig:Figure_1} presents the best fitting of {\tt GALFIT} for three typical examples 
of RS, GV and BC galaxies (see Section \ref{sec:galaxy classification} for classification), respectively. 
From left to right, panels in the first column show the raw images in F160W. 
Panels in the second and third columns show the disk and bulge components of two-component SerExp model, respectively. 
The black circles show the Kron apertures measured by {\tt SExtractor}, and the red circles indicate 
the 80\%-light radius of the bulge components ($R_{80,bulge}$), accounting for 80\% of the total bulge luminosities. 
Panels in the fouth column show the residual images after subtracting the two-component SerExp model 
from raw images. Panels in the last two columns show the growth curves of the bulge components 
and the comparisions between the observed and the best fitted radial profiles (PSF-convolved), respectively.

\section{Classification of Disk Galaxies} \label{sec:galaxy classification}

\begin{figure*}
\centering
\includegraphics[width=0.9\textwidth]{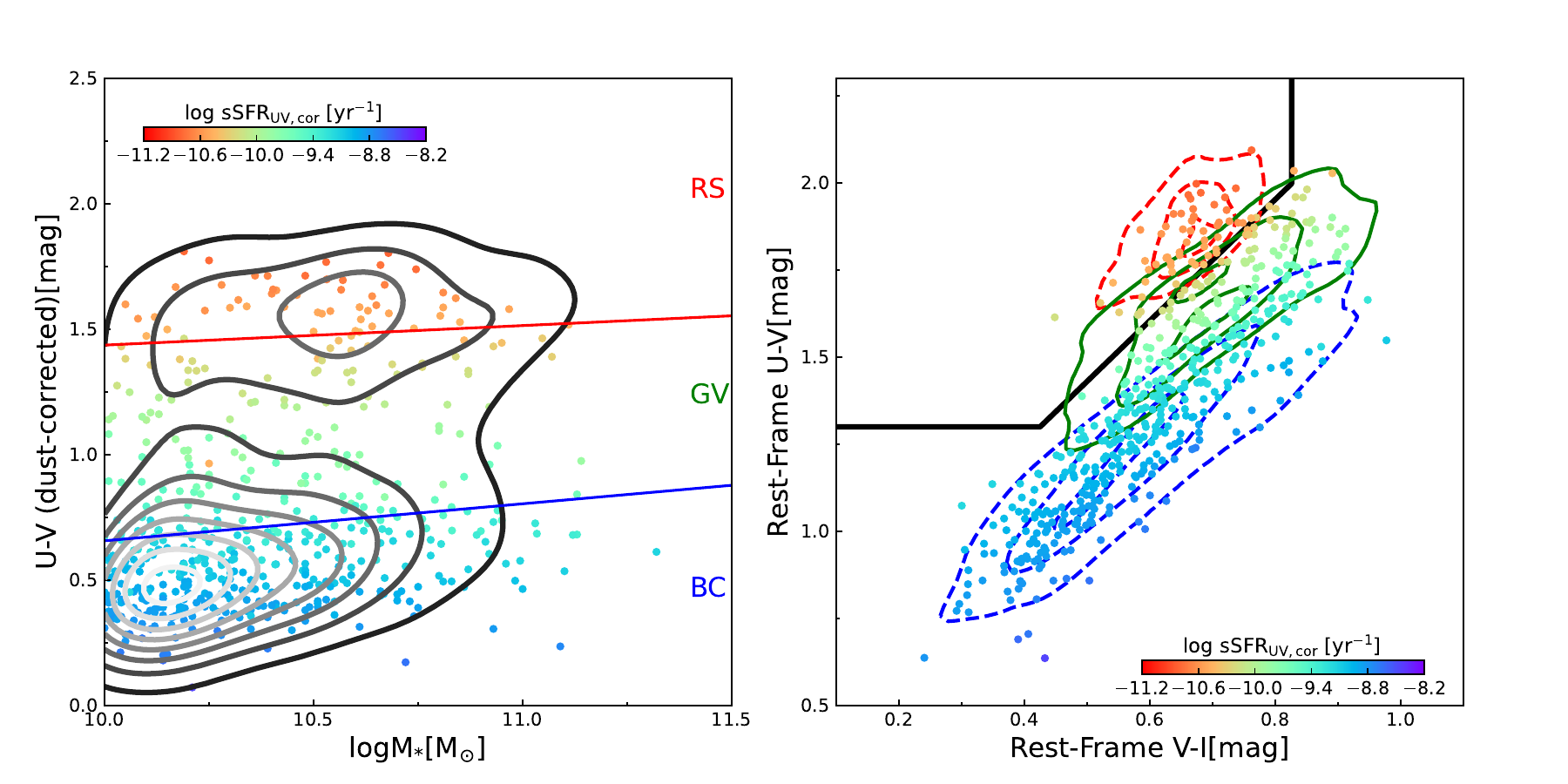}
\caption{\label{fig:Figure_2}
The left panel presents the dust-corrected rest-frame $U-V$ versus stellar mass diagram for disk galaxies.
The contours represent the relative densities of all massive galaxies with $M_{\ast} {\geq} 10^{10}M_{\odot}$
at $0.5 {\leq} z {\leq} 1.0$. 
The red and blue solid lines indicate the $2\sigma$ deviation below the best-fit RS 
and $2\sigma$ deviation above the best-fit BC, respectively. 
The right panel presents the distributions of RS, GV and BC galaxies in the
$UVI$ diagram, whose relative densities are shown by the red dash, green solid and blue dash contours, respectively.
The black solid lines indicate the classification criterion between star-forming and quiescent galaxies proposed
by \citet{Williams2009ApJ}. Data points in both panels are color-coded by $\rm log sSFR_{UV,cor}$.
}
\end{figure*}

Green valley galaxies represent an important population that is likely to transition from the 
blue cloud to the red sequence phases. It has been known that 
two-color diagrams (e.g., $UVJ$ or $UVI$) are not an effective tool to distinguish 
the GV from the RS and BC. In this work, 
we adopt the dust-corrected rest-frame $U-V$ color versus stellar mass diagram 
to divide our disk galaxies into RS, GV and BC sub-groups \citep[][]{Bell2004ApJ...608..752B}. 
Due to the different slopes of the color-stellar mass relations for RS and BC, we independently fit the best relations for RS and BC, and define GV galaxies as those that lie outside the 2-sigma distribution range of both populations.
To determine the best-fit RS, we first fit the relation for all $UVJ$-selected, 
massive quiescent galaxies, and then exclude objects that are more than 2$\sigma$ away from the fit 
and repeat the fitting process until no new objects are excluded. 
For the BC, we opt to fit the relation for galaxies with $\rm log sSFR_{UV,cor}>-9.5$, which 
are situated in close proximity to the star formation main sequence.
The derived best-fit RS is as follow,  
\[(U-V)_{rest}-{\Delta}A_V = 0.0779*log_{10}(M_*/M_{\odot})+0.8604\]
, and the best-fit BC is as follow,
\[(U-V)_{rest}-{\Delta}A_V = 0.1475*log_{10}(M_*/M_{\odot})-1.0304\]
where ${\Delta}A_V$ is equal to $0.49{\times}A_V$. 
The factor 0.49 is computed for the \citet{Calzetti2000ApJ} extinction law and the rest-frame Bessel filters. The best-fit RS and BC has a dispersion of 0.1013 and 0.1060 dex respectively.  

\begin{figure*}
\centering
\includegraphics[width=0.95\textwidth]{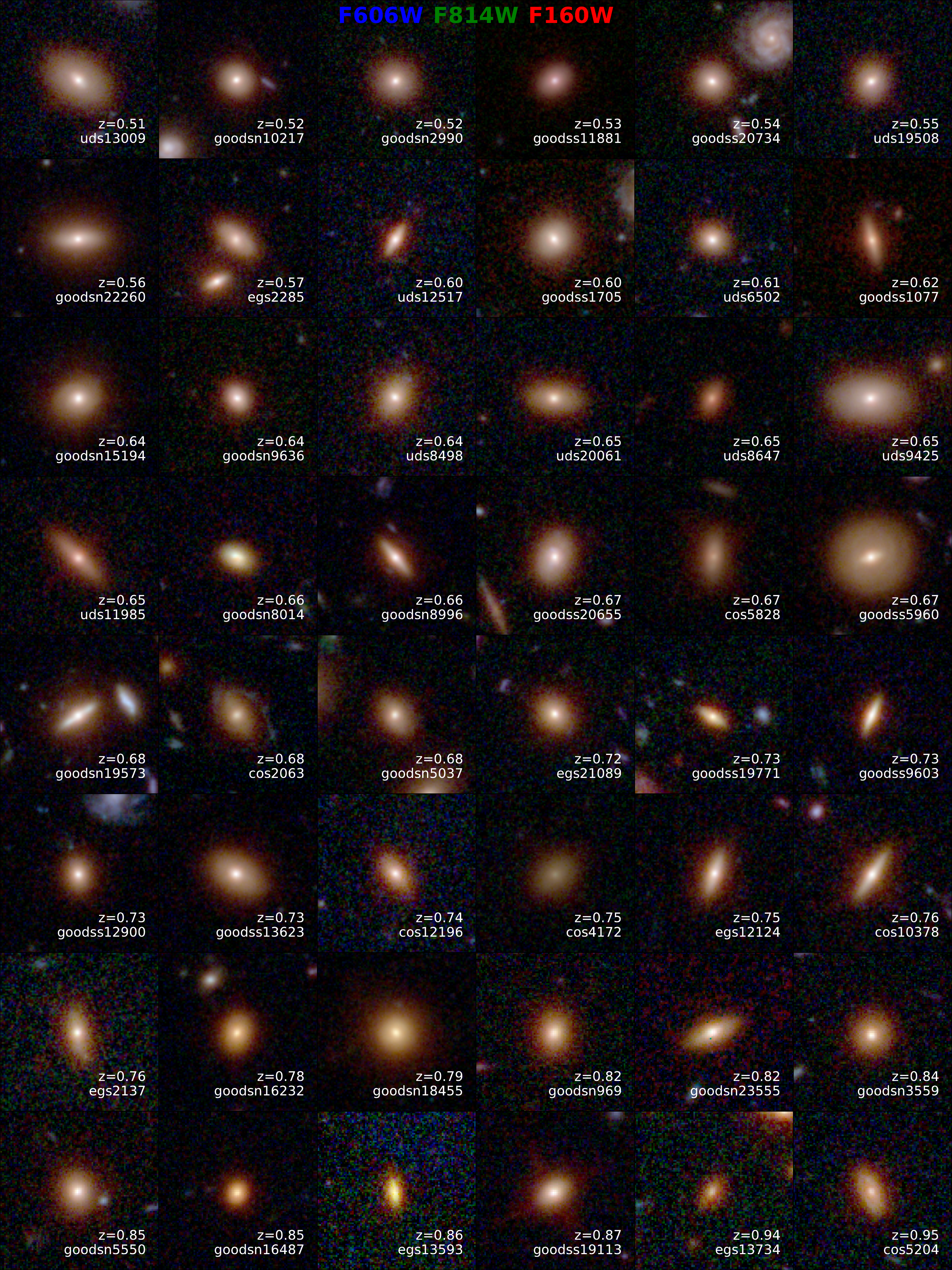}
\caption{\label{fig:Figure_3}Color thumbnail images of identified 48 disk-dominated RS galaxies. 
Each thumbnail has a size of $\rm 45 kpc \times 45 kpc$. The best redshift and CANDELS ID are displayed 
on the bottom-right conner of each thumbnail. 
}
\end{figure*}

To quantify the relative star formation activity in galaxies, 
we offset 2$\sigma$ below the best-fit RS and 2$\sigma$ above the best-fit BC, 
as shown by red and blue solid lines 
in the left panel of Figure \ref{fig:Figure_2}, respectively. Objects above the red line are classified 
as the RS, and those below the blue line are classified as the BC. Objects located in the region between 
the red and blue lines are defined as the GV. In the right panel of Figure \ref{fig:Figure_2}, 
we also show the distributions of RS, GV and BC galaxies in the $UVI$ diagram as a contrast. 
It can be seen that the RS, GV and BC galaxies are mixed partly in the two-color space. 
The figure illustrates that the dust-corrected color versus stellar mass diagram is indeed a 
more effective tool to identify the transition population, compared to the two-color diagram.  
Finally, we obtain 48 RS, 156 GV and 309 BC disk galaxies, respectively. 
The color thumbnail images of 48 disk-dominated RS galaxies are displayed in Figure \ref{fig:Figure_3}. 
The basic parameters of these 48 RS disk galaxies can be found in Table \ref{tab:table_2}. 

\section{Measurement of UVI Color Profiles} \label{sec:color profiles}

\begin{figure*}
\centering
\includegraphics[width=1.0\textwidth]{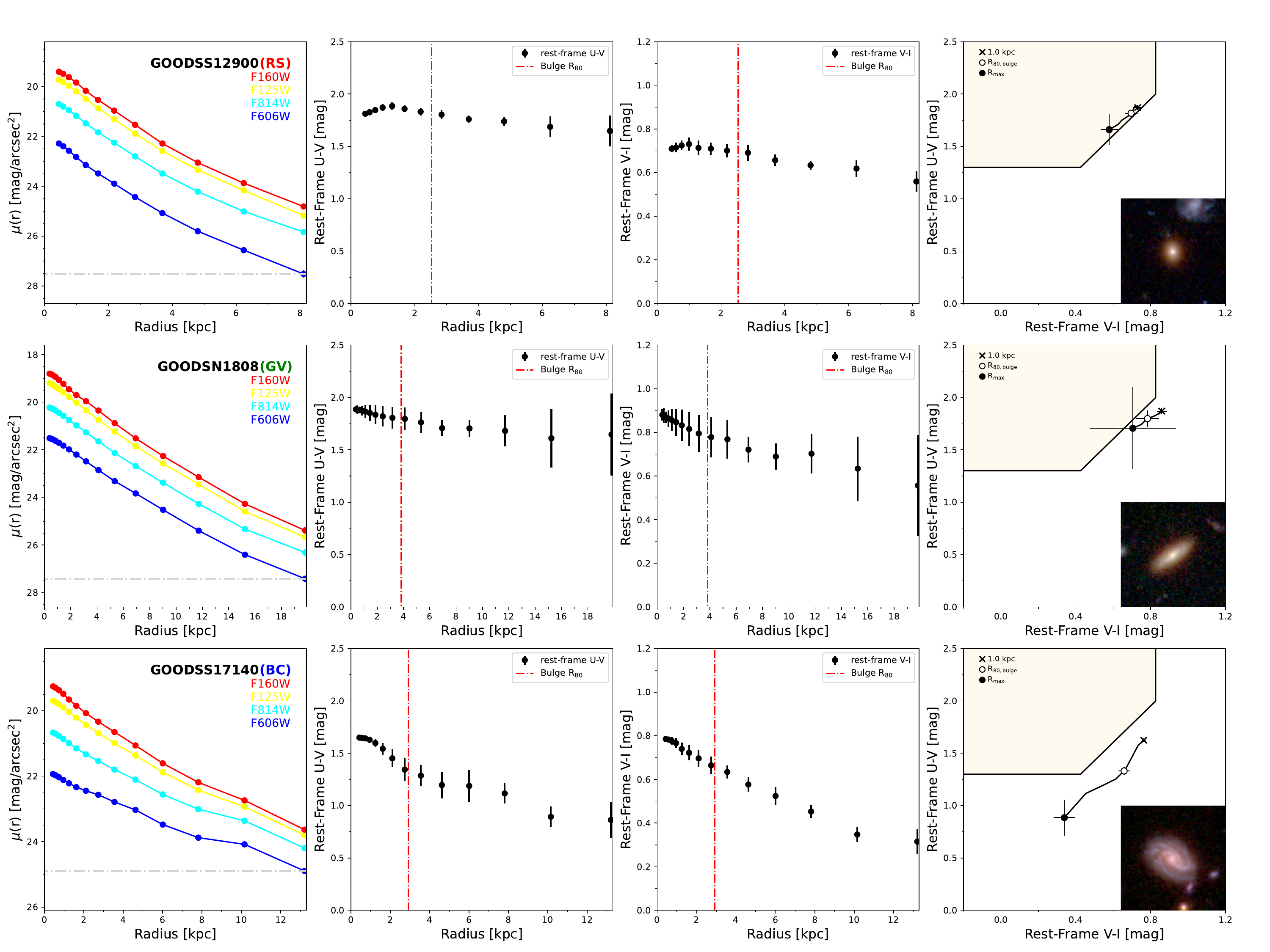}
\caption{\label{fig:Figure_4}
Panels in the left column show the observed surface brightness profiles in F606W, F814W, F125W 
and F160W bands for three typical examples of RS, GV and BC galaxies (from top to bottom), 
respectively. The horizontal dashed lines correspond to the $1\%$ of the local sky background level in F606W.
Panels in the middle two columns show their rest-frame $U-V$ and $V-I$ color profiles, 
respectively. The vertical dash-dotted line in each panel indicates the $R_{80,bulge}$. 
Panels in the right column show the $UVI$ color gradients in the $UVI$ diagram. 
The solid rectangles, empty circles and solid circles indicate the colors at 1 kpc, $R_{80,bulge}$ and 
maximum measured radius ($R_{max}$). The color thumbnail image is presented on the right bottom. 
} 
\end{figure*}


\begin{figure*}
\centering
\includegraphics[width=1.0\textwidth]{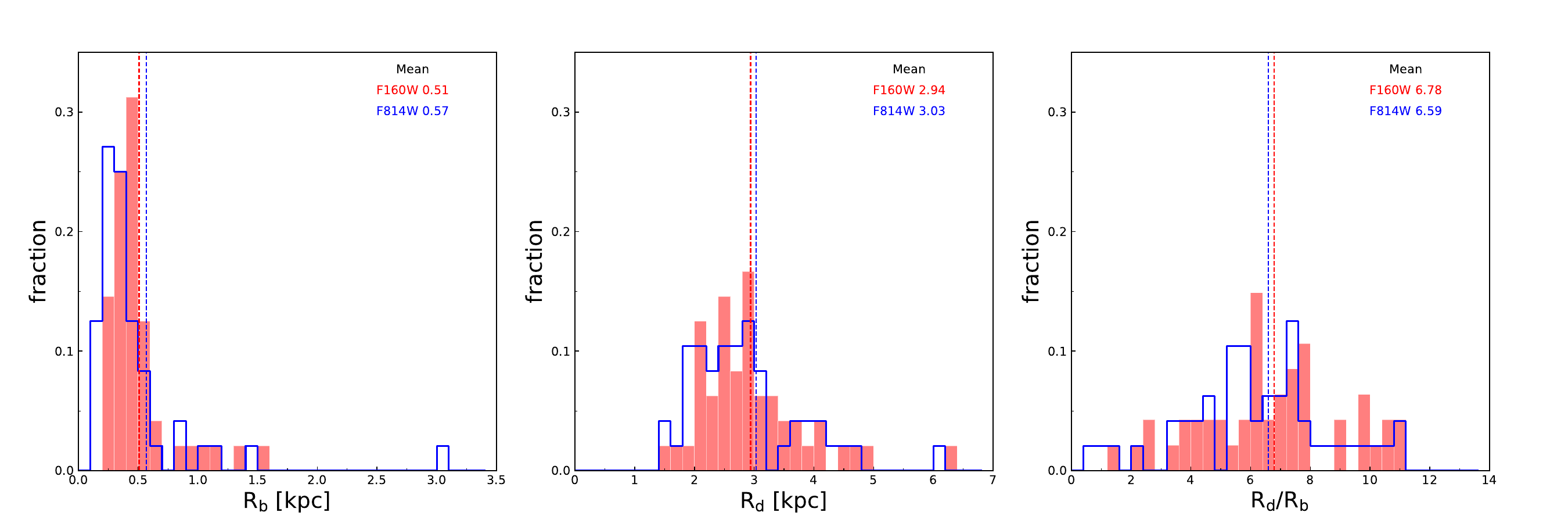}
\caption{\label{fig:Figure_5}
The distributions of bulge effective radii (left), disk effective radii (middle) 
and their ratios in the F814W (blue) and F160W bands for identified 48 RS disk galaxies, 
respectively. The mean values are marked with the vertical dashed lines.
}
\end{figure*}

We have measured the observed surface brightness profiles of sample galaxies 
in all WFC3/UVIS (F275W), ACS (F435W, F606W, F775W, F814W and F850LP) 
and WFC3/IR (F105W, F125W, F140W and F160W) bands, if available. 
The photometry was done by using the {\tt IRAF} routine {\tt ellipse}
within STSDAS, which is based on a technique described by \cite{Jedrzejewski1987}.
For the preliminary imaging reduction prior to multi-aperture photometry, we 
applied the same background-subtraction and noise-producing methods as described 
in Section \ref{sec:identificaiton of disks} to reduce all band images of individual galaxies. 
We then performed PSF-matching for the individual images of each band (except the F160W) 
to the F160W using the PSF matching kernels generated by {\tt photutils}.
We fixed the galaxy geometric centers, ellipticities and position angles 
obtained from the {\tt GALFIT} single-component S\'ersic fitting on F160W band imaging 
for all available bands.

It is noted that the two GOODS fields (GOODS-N and GOODS-S) have more HST high-resolution imaging 
data than other three fields (COSMOS, EGS and UDS). 
The images only in F606W, F814W, F125W and F160W are available simultaneously for all five 
CANDELS fields. 
In order to obtain consistent rest-frame UVI colors in different CANDELS fields, 
we determine the empirical relations between the observed colors combined 
among these four HST bands and rest-frame UVI colors for our galaxies. 
The empirical relations were constructed by fiting rich data in the GOODS-N and GOODS-S fields, 
which enables more accurate measurement of rest-frame colors based on the modeling 
of the spectral energy distributions (SEDs) of galaxies (see Appendix\ref{AppendixB.pdf} for details). 
In this analysis, we apply a uniform empirical relations for all galaxies 
within each narrow redshift bin. We convert the observed profiles in F606W, F814W, F125W and F160W 
to the rest-frame $U-V$ and $V-I$ color profiles for the sample galaxies in all five CANDELS fields.
We have verified that our principal results remain unchanged 
when distinct empirical relations are applied to galaxies with low- and high-sSFR, 
respectively. 
Among four observed bands, the F606W band has the lowest $S/Ns$ in the outskirts. We thus 
restrict the maximum acceptable radii (${R}_{max}$) for the profiles, where the surface brightness 
is equal to $1\%$ of the local sky background level in F606W. 
The mean ${R}_{max}/{R_e}$ values are 3.92, 3.48 and 2.85 for the RS, GV and BC, respectively.
Figure \ref{fig:Figure_4} illustrates our measurement procedure for three typical examples 
of RS, GV and BC disk galaxies, respectively. Panels in the left column show the surface brightness profiles 
in four observed bands (F606W, F814W, F125W and F160W). Panels in the middle two column show the 
rest-frame $U-V$ and $V-I$ profiles determined using the empirical relations 
in Appendix\ref{AppendixB.pdf}. Panels in the right column show the color profiles 
in the $UVI$ space, respectively.  

\section{Results and Analysis} \label{sec:results and analysis}

\subsection{Color gradients in RS disk galaxies} \label{red disks}

\begin{figure}
\centering
\includegraphics[width=0.48\textwidth]{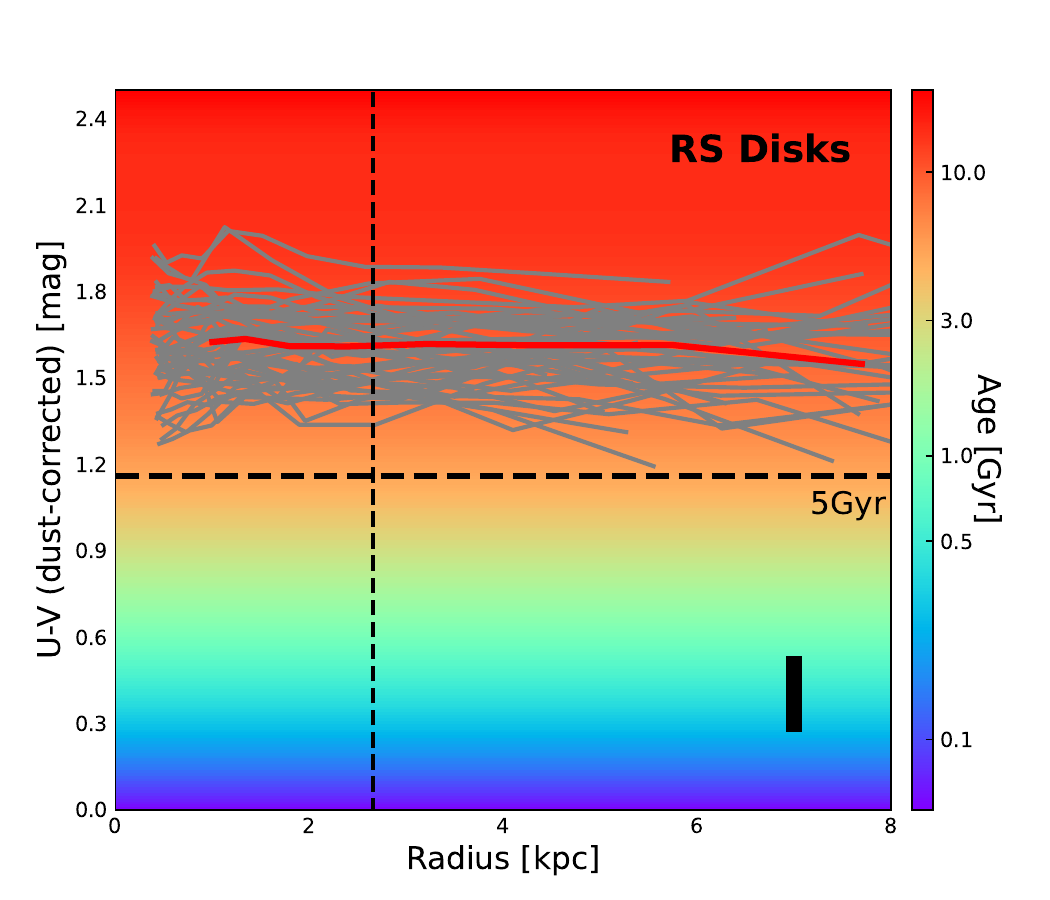}
\caption{\label{fig:Figure_6}
The dust-corrected $U-V$ radial profiles of identified 48 disk-dominated RS galaxies. 
The median profile is shown with the red solid line. 
The color-coded background show the age track for a dust-free, 
$\rm declining~\tau=3~Gyr$, solar-metallicity stellar population model from 
\citet{Bruzual2003MNRAS}. The age of 5 Gyr is indicated with a horizontal dashed line.
The regions outside the average $R_{80,bulge}$, indicated 
by the vertical dashed line, are where the disks dominate and the bulge components are negligible. 
The median uncertanity in the disk regions is presented on the bottom right corner.   
}
\end{figure}

We have identified a sample of 48 massive disk-dominated RS galaxies at $0.5\ {\leq}\ z\ {\leq}\ 1$ 
in CANDELS. 
In Figure \ref{fig:Figure_5}, we show the distributions of bulge effective radii, disk effective radii 
and their ratios in the F814W and F160W bands for these galaxies, respectively. 
It can be seen that these RS disk galaxies generally have extended stellar disks 
($\rm \sim 3~kpc$ on average) and tiny bulge components ($\rm \sim 0.5~kpc$ on average).
%
To quantify the intrinsic color gradients, we infer both the sSFR and $\rm Av$ profiles to
disentangle their effects on the raw dust-reddened $U-V$ color gradients of individual galaxies, 
applying the calibrations in Appendix\ref{Appendixa} to our spatially resolved data. 
In Figure \ref{fig:Figure_6}, we show the derived profiles of dust-corrected $U-V$ 
of individual RS disk galaxies. The median profile is presented in the red solid curve.  
In addition, we overplot the age track for a dust-free,
$\rm declining~\tau=3~Gyr$, solar-metallicity stellar population model from
\citet{Bruzual2003MNRAS}. 
It can be seen that these RS disk galaxies generally have intrinsically red colors and 
nearly flat $U-V$ color gradients. Both disk and bulge components have an old age.
%
The findings indicate that disk-donimated RS galaxies generally consist of uniformly red and old stellar populations 
in both the disk and bulge components, and the disks have generally been completely quenched.  

\subsection{Comparisons with the GV and BC counterparts} \label{quenching}

\begin{figure}
\centering
\includegraphics[width=0.48\textwidth]{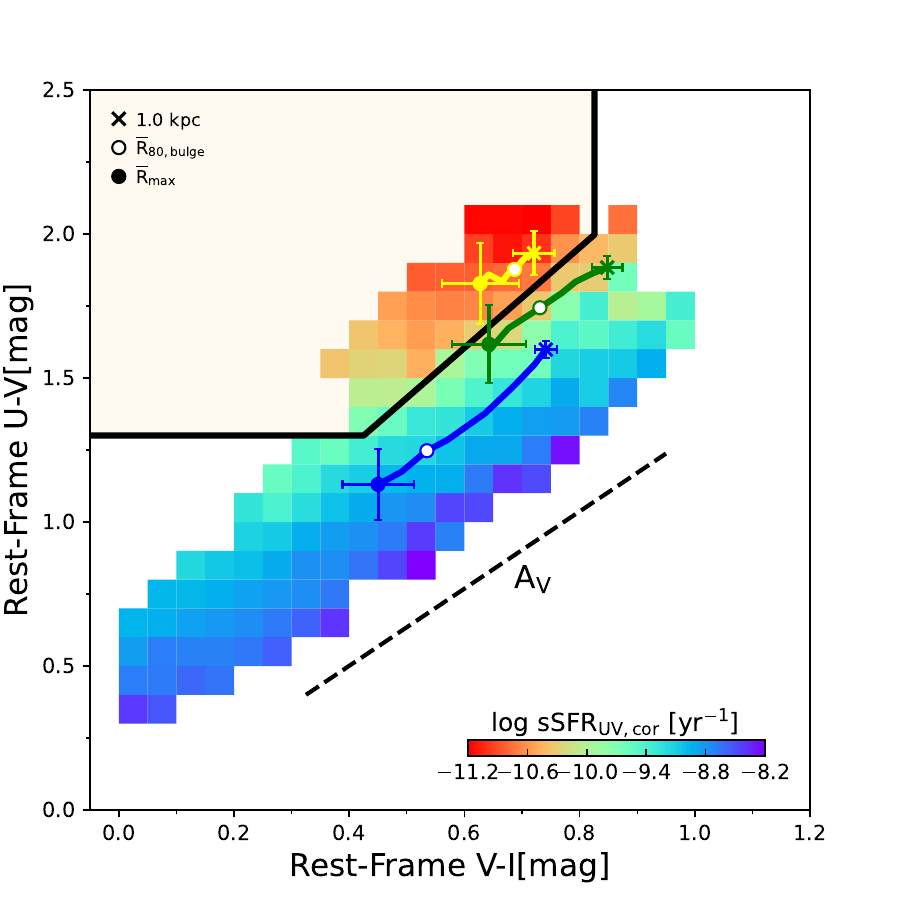}
\caption{\label{fig:Figure_7}
The stacked color trajectories in $UVI$-space for the identified
disk-dominated RS (yellow line), GV (green line) and BC (blue line) galaxies, respectively.
Three radial locations of 1 kpc, the average $R_{80,bulge}$ of the bulge components ($\overline R_{80,bulge}$)
and the average $R_{max}$ ($\overline R_{max}$) are marked by solid rectangles, empty circles and solid circles
for each sub-group, respectively.
}
\end{figure}

We have confirmed that on average the disk components in disk-dominated RS galaxies 
are truely quiescent and have nearly flat color gradients.
In consideration of their old stellar population and the intact disk morphology, 
it suggests that the star-forming progenitors of these galaxies 
probably consist of an extended disk and a small bulge (or nearly pure disk) as well, 
and the bulge growth should be negligible as they evolve. Otherwise, 
the buildup of larger, old bulges in the centers is likely to result in 
a structural transition from star-forming disks to quenched spheroids and/or 
larger color gradients. 
With the aim to understand this star formation quenching process, we make 
comparisions in the UVI color gradients and inferred sSFR gradients in different 
populations of disk-dominated galaxies (from the BC through the GV to the RS). 

In Figure \ref{fig:Figure_7}, we show raw stacked $UVI$ color trajectories 
in $UVI$-space for the RS, GV and BC disk galaxies, respectively. 
The stacked color profiles are computed by taking the median colors 
after applying the 3$\sigma$-clipping at different physical radii. Three radial 
locations of 1 kpc, the average $R_{80,bulge}$ of the bulge components ($\overline R_{80,bulge}$) 
and the average $R_{max}$ ($\overline R_{max}$) are marked for each sub-group, respectively. 
As seen from the figure, the median colors of RS disk galaxies at all radii lie well 
within the quiescent region, whereas the median colors of BC disk galaxies at all radii lie well 
within the star-forming region. The median colors of GV disk galaxies at all radii 
lie close to the boundary that separates quiescent and star-forming galaxies.  

\begin{figure*}
\centering
\includegraphics[width=1.0\textwidth]{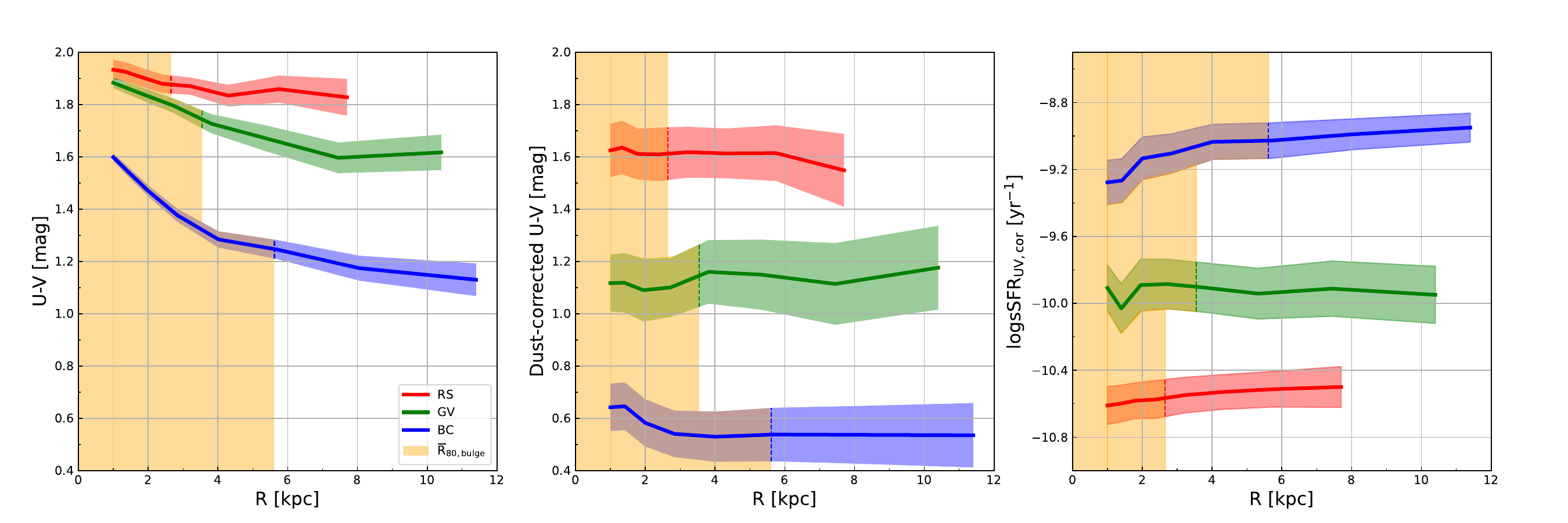}
\caption{\label{fig:Figure_8}
The stacked raw $U-V$ profiles (left), the dust-corrected $U-V$ profiles (middle) and 
the sSFR profiles (right) for the disk-dominated RS (red), GV (green)  and BC (blue) galaxies, respectively.
The orange shade regions indicate the average $R_{80,bulge}$ of the bulge components for each sub-group. 
}
\end{figure*}

Adopting the calibrations in Appendix\ref{Appendixa}, we convert $UVI$ olor trajectories to 
the stacked sSFR and $A_V$ profiles. 
In Figure \ref{fig:Figure_8}, we show the raw dust-reddened $U-V$ profiles, the dust-corrected $U-V$ 
profiles and the sSFR profiles for the RS, GV and BC disk galaxies, respectively. 
It can be seen that the BC disk galaxies exhibit larger, negative color gradients than 
both the GV and RS disk galaxies in the dust-reddened $U-V$ profiles, due to significant amounts 
of dust reddening. After correcting for dust reddening, we find that all the disk-dominated 
BC, GV and RS galaxies generally have nearly flat or weak 
gradients in both the dust-corrected $U-V$ profiles and the sSFR profiles.
%
%

\section{Discussion and Summary} \label{sec:Discussion and Conclusion}

A rare population of massive disk-dominated quiescent galaxies has recently drawn much attention,
which intrudes the RS without destroying the underlying stellar disks.
In this work, we have carefully identified a sample of 48 disk-dominated RS galaxies 
with $M_{\ast} > 10^{10}M_{\odot}$ between redshift 0.5 and 1.0 in all five CANDELS fields. 
These galaxies are well fitted by a two-component bulge plus disk model, 
and have the bulge-to-total ratio $B/T<0.4$ in the both F814W and F160W bands.
We show that these disk-dominated RS galaxies generally consist of uniformly red and old stellar populations
in both the disk and bulge components, and the disks have generally been completely quenched.
To understand their possible origins, we have also identifed the control samples of 124 GV and 
309 BC disk-dominated galaxies with the same selection criteria. 
We investigate the potential evolution of the stacked UVI color gradients and inferred sSFR gradients 
from the BC through the GV to the RS, out to the maximum acceptable radii (${R}_{max}$), 
where the surface brightness is equal to $1\%$ of the local sky background level 
in F606W (the average values, $\overline{R}_{max}$, are $\sim7.7$ kpc, $\sim10.4$ kpc and $\sim$11.4 kpc 
for the RS, GV and BC disks, respectively).
After correcting for dust reddening, we find that all disk-dominated RS, GV and BC galaxies 
generally have nearly flat $U-V$ color profiles and sSFR profiles. 
We propose a potential formation mechanism that massive quiescent disk-dominated galaxies 
at moderate redshifts are predominantly formed via a process of secular disk fading.
%

Regarding the PSF-smearing effect, uncertainty of PSF matching, and error estimate 
in the measurement of color gradients, these have all been discussed in \citet{Liu+18} and 
\citet[][]{Wang2017MNRAS}. We do not duplicate the descriptions here, and refer readers to these two 
papers for details. 
We have checked that our main results remain unchanged under the stacking by 
scaled radius (i.e., normalized by effective radius) rather than physical radius. 
The aim of displaying profiles in physical radius is to highlight a phenomenon that 
quenching is associated with a shrinkage in optical size, which is in agreement with 
previous studies \citep[][]{Pandya2017,Fang2018ApJ}.   
The overall nearly flat sSFR profiles in the disk-dominated BC, GV and RS galaxies appear to
be in tension with the shinkage in size. 
One potential explanation for this discrepancy is that the half-mass radii of star-forming galaxies are generally much smaller than their half-light radii, which has been confirmed 
in several studies \citep[e.g.,][]{2013ApJ...763...73S,2019ApJ...885L..22S,2023ApJ...945..155M}.
As previously noted, the average ratios of ${R}_{max}/{R_e}$ for our BC, GV, and RS disk galaxies 
are 2.85, 3.48, and 3.92, respectively. We exercise caution in acknowledging that we may be probing distinct regions within the three subcategories of galaxies. This could be another interpretation for the observed 
discrepancy. Future deeper observational images will aid in ascertaining whether these galaxies exhibit significant color gradients at larger radii.

\begin{figure*}
\centering
\includegraphics[width=0.92\textwidth]{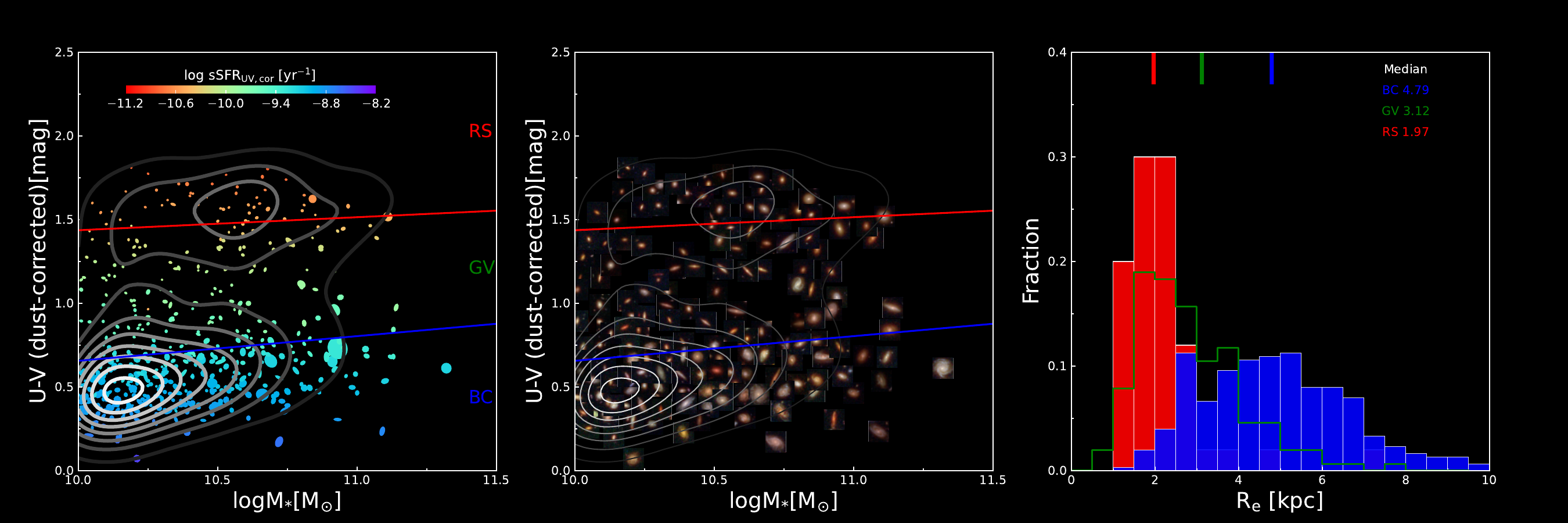}
\caption{\label{fig:Figure_9}
Left: The distribution of identified disk-dominated galaxies in the dust-corrected $U-V$ 
color versus stellar mass diagram, coded by the structural parameters (the normalized effective radii, 
axis ratios and position angles). The data points are also color-coded 
by $\rm log sSFR_{UV,cor}$ simultaneously. 
Middle: Same as the left panel, but shown with their color thumbnail images with 
a nonoverlapping foreground layer. 
Right: The distributions of effective radii from single-S\'ersic fits for RS, GV, and BC disk galaxies, 
respectively. 
The vertial lines on the top indicates the median values for different populations, which 
are shown on the top right.    
}
\end{figure*}

GV galaxies are thought to be in the transition phase of galaxy evolution. 
%
In order to further understand this phase, we also investigate the similarites in structure between 
the disk-donimated GV and RS galaxies. 
In the left panel and middle panel of Figure \ref{fig:Figure_9}, we again show the distributions 
of identified disk-dominated galaxies in the space of dust-corrected $U-V$ color versus stellar mass, 
but this time galaxies are coded by their structural parameters (i.e., effective radii, axis ratios 
and postion angles), and are shown with their color thumbnail images, respectively. 
We specially show the distributions of their effective radii 
in the right panel of Figure \ref{fig:Figure_9}.
As seen from these plots, the disk-dominated GV galaxies appear to have more similar structures 
to the RS than to the BC. This again supports that they are likely to be 
transitioning to form disk-dominated RS galaxies.
This work is focus on the epoch of $z=0.5-1$, which coresponds to an age interval of 
$\rm \sim2.7~Gyr$. At this epoch, quenching is expected to be a slow process, related 
with gas consumption in the star-forming disks \citep[][]{Fang2013ApJ...776...63F,Barro2017ApJ...840...47B}. 
However, we still caution that some disk-dominated RS galaxies may form directly 
through highly dissipative events with significant angular monmentum 
at higher redshifts \citep[][]{McGrath2008ApJ,Fudamoto_2022} , 
although quenching at higher redshifts is usually triggered by a compaction 
event that builds up a compact core \citep[][]{Barro2014,Barro2017ApJ...840...47B}.  

It is noted that quiescent disks are likely a collection of several individual populations 
that are a product of different quenching mechanisms and effects. It is still far from well understood 
how massive quiescent disks formed, and how the star formaiton was quenched with intact disk morphology. 
This study is focused on the properties of disk-dominated RS galaxies and 
the probable origins of quiescent disks within them, excluding those RS disk galaxies dominated by bulges.
We propose the possibility that the quenched disks in these galaxies could form through a process of secular disk fading. It has been observed that bulge-dominated RS disk galaxies are most likely to possess the substructures of strong bars and spirals \citep[e.g.,][]{GuoRui2020ApJ...897..162G,Bundy2010ApJ...719.1969B}. 
Consequently, the cessation of star formation in bulge-dominated disk galaxies is likely to be strongly correlated with the bar instability and/or the development of large bulges.
In contrast, it can be seen in Figure 3 that the disk-dominated RS galaxies generally exhibit relatively 
smooth morphology, lacking the features of distinct bars and spirals. 
Additionally, it is still observable in the middle panel of 
Figure \ref{fig:Figure_8} that there appears to be a negative color gradient as one transitions from the bulge to the disk in RS disk galaxies, with BC galaxies also exhibiting a slight gradient. 
This may suggest that the cessation of star formation in disk-dominated galaxies 
could also be marginally correlated with the growth of the bulge. 
Therefore, our current results only support the idea of strangulation or starvation, acknowledging that 
other mechanisms (e.g., central compaction and AGN feedback) might contribute, 
as an important part of the mechanisms that creates quiescent disks. 
In a subsequent work, we will investigate the properties 
of quiescent disks in bulge-dominated RS galaxies, and discuss various possible formation 
mechanisms with a spectroscopic sample including the JWST+HST multi-band imaging observations.

~\\\\
We thank anonymous referee for the insightful suggestions, which significantly helped us improve this paper.
We thank Drs. Jian Ren, Man I. Lam, Jiao Li, Xin Zhang, Xianmin Meng and Juanjuan Ren for useful suggestions and discussions. 
This project is supported by the National Natural Science Foundation of China (NSFC grants Nos. 12273052, 11733006, 12090040, 12090041 and 12073051) and the science research grants from the China Manned Space Project (No. CMS-CSST-2021-A04).

\begin{appendix}
\section{Recalibrations of Dust Extinction and specific Star Formation Rates} \label{Appendixa}

\begin{figure*}
\centering
\includegraphics[width=0.75\textwidth]{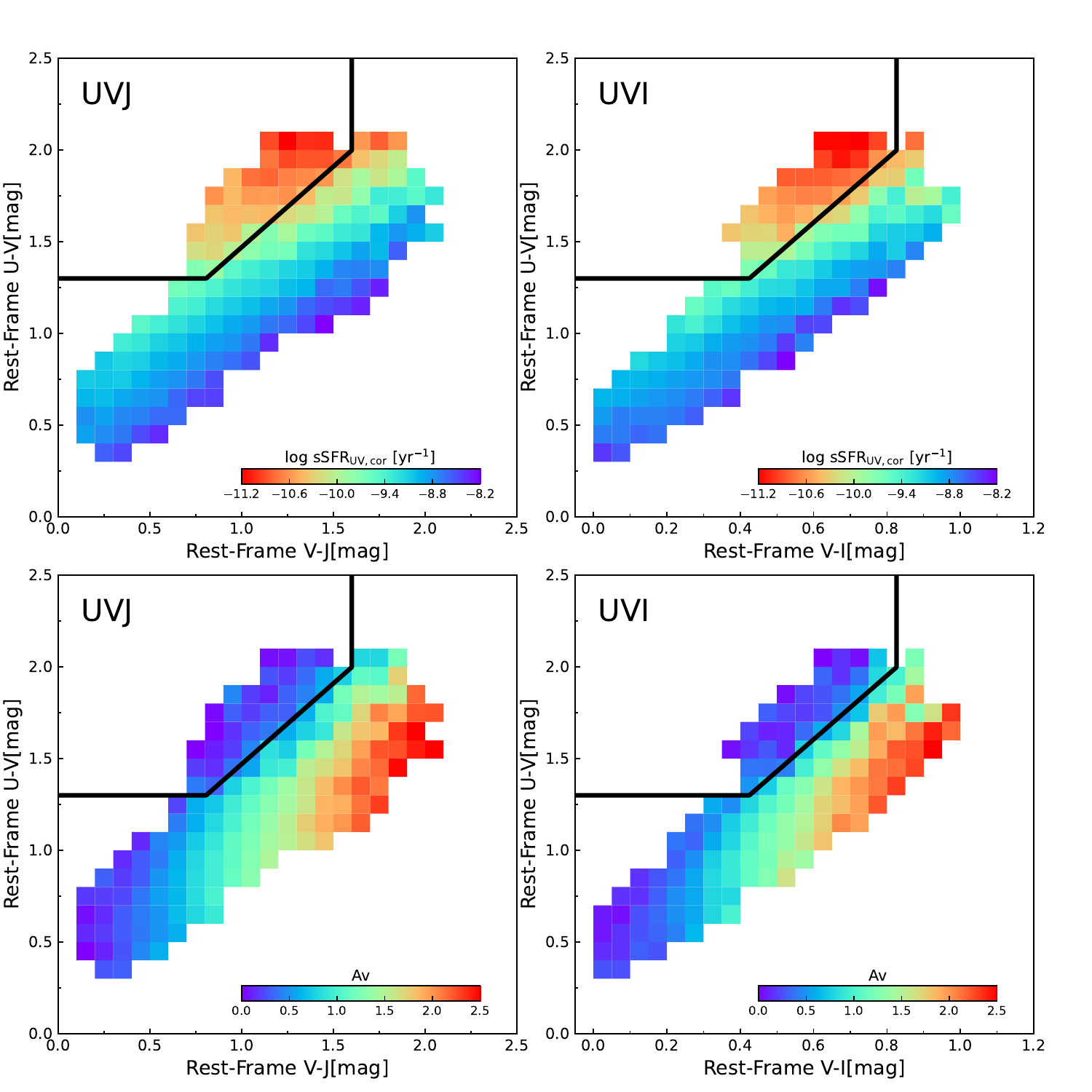}
\caption{\label{fig:AppendixA}
Two-color diagrams for CANDELS galaxies with $M_* {\geq} 10^{9}M_{\odot}$ at $0.5 {\leq} z {\leq} 1.0$. 
Left: $UVJ$ diagrams color-coded by the median $\rm log sSFR_{UV,cor}$ (top) and 
$\rm A_V$ in classified boxes. 
The black lines serve as the boundary to separate quiescent galaxies from star-forming galaxies 
following \citet{Williams2009ApJ}. Right: the corresonding $UVI$ diagrams color-coded 
by the median $\rm log sSFR_{UV,cor}$ (top) and $\rm A_V$ in classified boxes, respectively. 
The black lines serve as the boundary to separate quiescent galaxies from star-forming galaxies 
foloowing \citet{Wang2017MNRAS}. Each box contains more than three galaxies. 
}
\end{figure*}

Rest-frame $UVJ$ diagram has been widely used to separate quiescent from dusty/star-forming galaxies 
\citep[e.g.][]{Williams2009ApJ,Fang2018ApJ}. It has also been successfully utilized to determine $\rm A_V$ and 
sSFR values, which are broadly consistent with the values derived from fitting reddened stellar population 
models to broadband SEDs of galaxies covering UV to Infrared \citep[][]{Fang2018ApJ}. 
Based on the data in two of the five CANDELS fields (GOODS-S and UDS), \citet[][]{Wang2017MNRAS} 
demonstrated that the rest-frame $UVI$ diagram is as useful 
as the $UVJ$ diagram for distinguishing $\rm A_V$ from sSFR, by replacing $J$ with a slightly bluer $I$ 
band. \citet[][]{Wang2017MNRAS} showed that sSFR and $\rm A_V$ can be constrained using 
$UVI$ with an $rms$ accuracy of 0.15 dex and 0.18 mag, respectively.   

In this work, we revisit the calibrations of $\rm A_V$ and sSFR using $UVI$ with enlarged sample of 
galaxies in CANDELS. In Figure \ref{fig:AppendixA}, we show the distribution of a carefully-selected subsample 
of galaxies on the $UVI$ plane, as well as on $UVJ$ for a comparison. 
Galaxies are selected after applying the cuts 1-3 with $M_* {\geq} 10^9M_{\odot}$ at $0.5 {\leq} z {\leq} 1.0$ 
in four of the five CANDELS fields (EGS, GOODS-N, GOODS-S and UDS). 
We find that the quality of CANDELS/COSMOS official photometry catalog is not as high as 
those of other four fields, and including it would not improve the accuracy of calibrations. 
We thus excluded the COSMOS data for the calibrations. 
As seen from the figure, the $UVI$ reproduces all the main features of $UVJ$, including 
the quiescent region and the distinctive stripe patterns of sSFR and $\rm A_V$, which is 
in agreement with the results initially presented by \citet[][]{Wang2017MNRAS}. 
We divide each plane into 25 $\times$ 25 rectangular boxes and compute the median values 
of $A_V$ and sSFR in each box, assigning them to the box center. 
We then interpolate linearly among the nearest box centers to determine the corresponding $A_V$ and sSFR 
for any given location on the $UVI$ plane. As a result, this method enables us to estimate 
sSFR and $A_V$ using $UVI$ with an rms accuracy of $\sim0.12$ dex and $\sim0.13$ mag, which is slightly 
higher than that of \citet[][]{Wang2017MNRAS}.

\section{Inprovements on Empirical Rest-frame UVI Colors} \label{AppendixB.pdf}

\begin{figure*}
\centering
\includegraphics[width=1.0\textwidth]{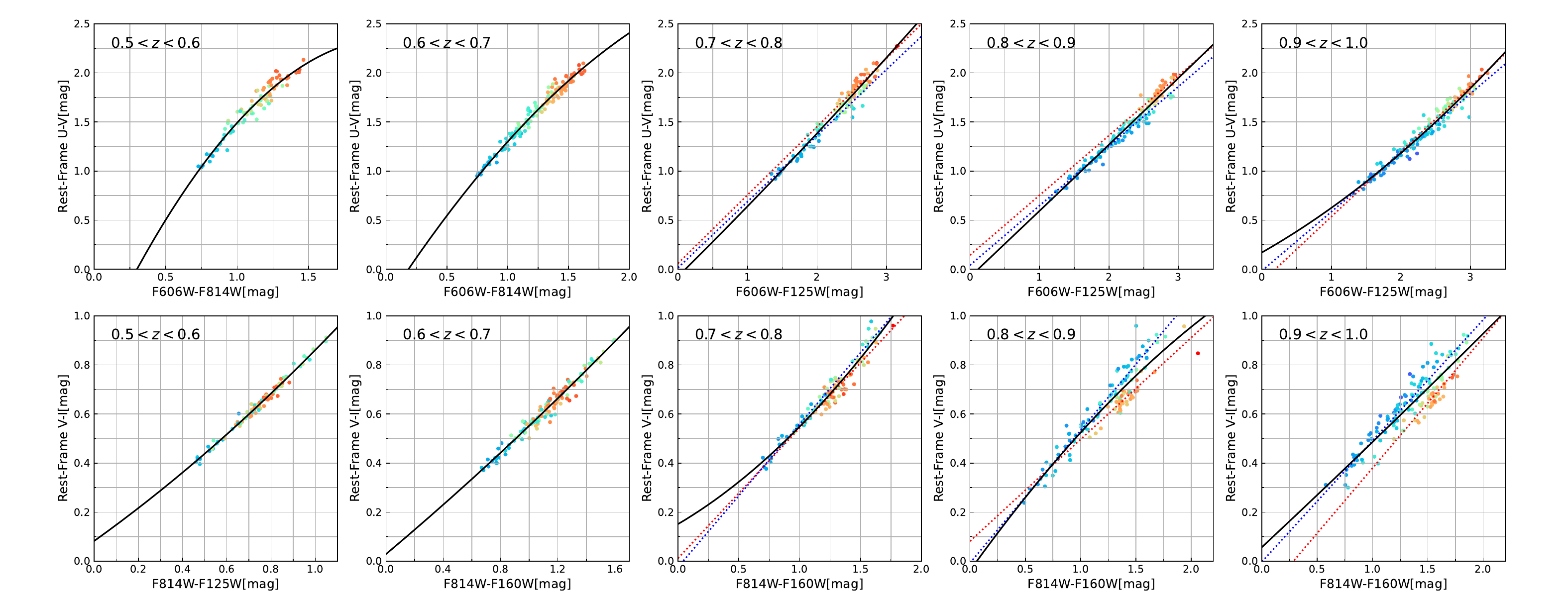}
\caption{\label{fig:AppendixB} Fitting the relations between 
the rest-frame $UVI$ colors and various observer-frame filter combinations 
for CANDELS massive galaxies with $M_* {\geq} 10^{10}M_{\odot}$ within each narrow redshift interval. 
Each point in the plots represents the integrated colors of one galaxy in the GOODS-N and GOODS-S fields. 
Data points are color-coded by $\rm log sSFR_{UV,cor}$. 
The solid curves represent the best single-form fit for all galaxies in each redshift bin. 
For three high-redshift bins ($z=0.7-1$), the best fits 
for the low-sSFR (red) and high-sSFR (blue) galaxies are presented with the dotted lines, separately.
%
}
\end{figure*}

It should be noted that the two GOODS fields (GOODS-N and GOODS-S) have more HST high-resolution imaging data than other three fields (COSMOS, EGS and UDS), in which only $F606W(V)$, $F814W(I)$, $F125W(J)$ and $F160W(H)$ photometry is available. 
In order to derive rest-frame color profiles using multi-band aperture photometry and obtain consistent results for all five CANDELS fields, it is necessary to determine the empirical relations between rest-frame colors and observer-frame colors from the integrated photometry.
\citet[][]{Wang2017MNRAS} have shown the robustness of rest-frame $UVI$ colors derived from $VIJH$, and that there are no systematic difference between the rest-frame colors derived from $VIJH$ bands and those derived from the seven $BVizYJH$ bands. 
However, the empirical relations derived by \citet[][]{Wang2017MNRAS} have relatively large scatters (see their Figure A1 and A2), partly due to the use of a bit broad redshift bins and all the five fields. 
In order to improve these empirical relations, we only exploit the GOODS-N and GOODS-S fields, which are more data-rich and enable more accurate measurement of rest-frame colors based on the modelingof the SEDs of galaxies.
We measured fluxes in 10 bands (F275W, F435W, F606W, F775W, F814W, F850LP, F105W, F125W, F140W 
and F160W) for all galaxies obtained through cuts 1-9. Subsequently, we fit their SEDs 
with a fixed redshift to derive the rest-frame $UVI$ colors.
We first perform a uniform fit for all galaxies within each narrow redshift bin. 
The fitting results are presented with solid lines in Figure \ref{fig:AppendixB}. 
The typical scatter of our fittings is $\sim 0.04$ mag for rest-frame $U-V$ and $\sim 0.02$ mag for rest-frame $V-I$, respectively. 
The fitting equations for each narrow redshift bin are given as follows:  

For $0.5<z<0.6$:
\[Restframe\ U-V = -0.7481{\times}(F606W-F814W)^2+3.1037{\times}(F606W-F814W)-0.8625\]
\[Restframe\ V-I = 0.1339{\times}(F814W-F125W)^2+0.6448{\times}(F814W-F125W)+0.0820\]

For $0.6<z<0.7$:
\[Restframe\ U-V = -0.2676{\times}(F606W-F814W)^2+1.9101{\times}(F606W-F814W)-0.3423\]
\[Restframe\ V-I = 0.0320{\times}(F814W-F160W)^2+0.4907{\times}(F814W-F160W)+0.0290\]

For $0.7<z<0.8$:
\[Restframe\ U-V = 0.0146{\times}(F606W-F125W)^2+0.6998{\times}(F606W-F125W)-0.0734\]
\[Restframe\ V-I = 0.1066{\times}(F814W-F160W)^2+0.2908{\times}(F814W-F160W)+0.1512\]

For $0.8<z<0.9$:
\[Restframe\ U-V = 0.0017{\times}(F606W-F125W)^2+0.6708{\times}(F606W-F125W)-0.0782\]
\[Restframe\ V-I = -0.0659{\times}(F814W-F160W)^2+0.6282{\times}(F814W-F160W)-0.0382\]

For $0.9<z<1.0$:
\[Restframe\ U-V = 0.0514{\times}(F606W-F125W)^2+0.4037{\times}(F606W-F125W)+0.1709\]
\[Restframe\ V-I = 0.0066{\times}(F814W-F160W)^2+0.4219{\times}(F814W-F160W)+0.0573\]

Although our fitted relations exhibit less scatter than those of \citet[][]{Wang2017MNRAS}, 
it is still observed that certain features persist in the high redshift bins (z=0.7-1). 
Galaxies with low sSFR appear to follow slightly different relations compared to those with high sSFR. 
To estimate the potential impact of this difference, we additionally 
calculate the median  $\rm log sSFR_{UV,cor}$ of galaxies at $z=0.7-1$ to 
divide the sample galaxies into low-sSFR and high-sSFR subgroups. 
We present the best fits for the low-sSFR and high-sSFR galaxies separately, 
as indicated by the dotted lines in Figure \ref{fig:AppendixB}. 
The fitting equations for low-sSFR and high-sSFR galaxies in the three high-redshift bins are as follows:
%
%

For $0.7<z<0.8$:

low-sSFR:
\[Restframe\ U-V = 0.6966{\times}(F606W-F125W)+0.0601\]
\[Restframe\ V-I = 0.5309{\times}(F814W-F160W)+0.0119\]

high-sSFR:
\[Restframe\ U-V = 0.6736{\times}(F606W-F125W)+0.0146\]
\[Restframe\ V-I = 0.5826{\times}(F814W-F160W)-0.0243\]

For $0.8<z<0.9$:

low-sSFR:
\[Restframe\ U-V = 0.6097{\times}(F606W-F125W)+0.1439\]
\[Restframe\ V-I = 0.4145{\times}(F814W-F160W)+0.0818\]

high-sSFR:
\[Restframe\ U-V = 0.6072{\times}(F606W-F125W)+0.0409\]
\[Restframe\ V-I = 0.5401{\times}(F814W-F160W)-0.0074\]

For $0.9<z<1.0$:

low-sSFR:
\[Restframe\ U-V = 0.6636{\times}(F606W-F125W)-0.1270\]
\[Restframe\ V-I = 0.5307{\times}(F814W-F160W)-0.1510\]

high-sSFR:
\[Restframe\ U-V = 0.6019{\times}(F606W-F125W)-0.0154\]
\[Restframe\ V-I = 0.4931{\times}(F814W-F160W)-0.0021\]

\end{appendix}

\bibliography{sample631}{}
\bibliographystyle{aasjournal}

\setlength{\tabcolsep}{1pt}
\begin{sidewaystable}[hbpt]
	\tiny
	\centering
	\caption{Basic parameters of the identified 48 RS disk galaxies in this work \label{tab:table_2}}
	\begin{tabular}{cccccccccccccccccccccccc}
		\hline
		\colhead{Field} & \colhead{ID} & \colhead{RA} & \colhead{DEC} & \colhead{zbest} & \colhead{$\rm logM_*$} & \colhead{$\rm m_{D814}$} & \colhead{$\rm Re_{D814}$} & \colhead{$\rm b/a_{D814}$} & \colhead{$\rm PA_{D814}$} & \colhead{$\rm m_{B814}$} & \colhead{$\rm Re_{B814}$} & \colhead{$\rm b/a_{B814}$} & \colhead{$\rm PA_{B814}$} & \colhead{$\rm n_{B814}$} & \colhead{$\rm m_{D160}$} & \colhead{$\rm Re_{D160}$} & \colhead{$\rm b/a_{D160}$} & \colhead{$\rm PA_{D160}$} & \colhead{$\rm m_{B160}$} & \colhead{$\rm Re_{B160}$} & \colhead{$\rm b/a_{B160}$} & \colhead{$\rm PA_{B160}$} & \colhead{$\rm n_{B160}$} \\
		\colhead{} & \colhead{} & \colhead{(degree)} & \colhead{(degree)} & \colhead{} & \colhead{($\rm M_{\odot}$)} & \colhead{(mag)} & \colhead{(arcsec)} & \colhead{} & \colhead{(degree)}  & \colhead{} & \colhead{(mag)} & \colhead{(arcsec)} & \colhead{(degree)} & \colhead{} & \colhead{(mag)} & \colhead{(arcsec)} & \colhead{} & \colhead{(degree)} & \colhead{(mag)} & \colhead{(arcsec)} & \colhead{} & \colhead{(degree)} & \colhead{} \\
		\hline
		COSMOS & 2063 & 150.092551 & 2.203012 & 0.68 & 10.64 & 21.99 & 0.58 & 0.55 & 30.35 & 22.67 & 0.09 & 0.66 & -50.92 & 1.21 & 20.79 & 0.50 & 0.57 & 31.39 & 21.27 & 0.08 & 0.67 & -56.08 & 1.11 \\
		COSMOS & 4172 & 150.060340 & 2.225983 & 0.75 & 10.74 & 21.58 & 0.43 & 0.73 & -50.24 & 23.82 & 0.05 & 0.58 & 70.20 & 1.80 & 20.36 & 0.42 & 0.72 & -48.96 & 22.47 & 0.04 & 0.50 & 64.30 & 2.81 \\
		COSMOS & 5204 & 150.145050 & 2.239186 & 0.95 & 10.64 & 22.45 & 0.53 & 0.47 & 19.51 & 23.98 & 0.14 & 0.59 & 15.72 & 8.00 & 20.92 & 0.43 & 0.47 & 21.28 & 23.09 & 0.03 & 0.12 & 26.09 & 4.47 \\
		COSMOS & 5828 & 150.195351 & 2.246359 & 0.67 & 10.81 & 21.14 & 0.42 & 0.40 & -4.41 & 22.75 & 0.09 & 0.68 & -75.40 & 8.00 & 19.91 & 0.37 & 0.45 & -4.98 & 21.71 & 0.05 & 0.58 & -84.28 & 0.50 \\
		COSMOS & 10378 & 150.118156 & 2.302016 & 0.76 & 10.81 & 21.59 & 0.45 & 0.21 & -32.35 & 22.42 & 0.07 & 0.62 & -27.94 & 0.85 & 20.41 & 0.44 & 0.22 & -33.21 & 21.30 & 0.06 & 0.87 & -46.79 & 0.50 \\
		COSMOS & 12196 & 150.062499 & 2.323891 & 0.74 & 10.54 & 22.43 & 0.43 & 0.54 & 37.19 & 23.11 & 0.05 & 0.52 & 47.61 & 2.36 & 21.08 & 0.40 & 0.54 & 36.82 & 21.84 & 0.06 & 0.53 & 58.68 & 1.33 \\
		EGS & 2137 & 214.824035 & 52.734187 & 0.76 & 10.65 & 22.24 & 0.58 & 0.41 & 9.33 & 23.00 & 0.05 & 0.71 & -19.81 & 1.13 & 20.98 & 0.54 & 0.40 & 11.07 & 21.46 & 0.05 & 0.60 & -30.41 & 2.58 \\
		EGS & 2285 & 215.055783 & 52.898627 & 0.57 & 10.50 & 21.48 & 0.45 & 0.58 & 49.75 & 23.56 & 0.08 & 0.55 & -14.03 & 0.55 & 20.68 & 0.42 & 0.54 & 48.72 & 21.21 & 0.18 & 0.61 & 43.81 & 1.68 \\
		EGS & 12124 & 215.143248 & 53.015315 & 0.75 & 10.67 & 22.08 & 0.43 & 0.28 & -17.35 & 22.85 & 0.06 & 0.64 & -32.25 & 6.13 & 20.71 & 0.39 & 0.30 & -17.81 & 21.57 & 0.04 & 0.60 & -14.59 & 8.00 \\
		EGS & 13593 & 214.772904 & 52.762324 & 0.86 & 10.10 & 23.82 & 0.33 & 0.37 & 6.90 & 25.98 & 0.12 & 0.24 & -0.55 & 0.50 & 22.55 & 0.37 & 0.42 & 5.35 & 23.16 & 0.18 & 0.21 & 8.90 & 0.77 \\
		EGS & 13734 & 214.677786 & 52.695180 & 0.94 & 10.40 & 23.97 & 0.39 & 0.47 & -34.04 & 25.25 & 0.06 & 0.08 & -28.26 & 0.92 & 22.19 & 0.31 & 0.51 & -35.05 & 23.59 & 0.09 & 0.18 & -14.74 & 0.50 \\
		EGS & 21089 & 215.153549 & 53.069605 & 0.72 & 10.43 & 22.21 & 0.36 & 42.88 & 0.79 & 23.03 & 0.05 & 0.67 & 24.79 & 0.50 & 21.00 & 0.36 & 0.80 & 38.44 & 21.64 & 0.05 & 0.75 & 52.44 & 0.50 \\
		GOODSN & 969 & 189.203040 & 62.128785 & 0.82 & 10.74 & 22.32 & 0.36 & 0.74 & -4.76 & 23.33 & 0.09 & 0.62 & -41.85 & 0.95 & 20.90 & 0.37 & 0.71 & -6.02 & 21.61 & 0.08 & 0.69 & -57.03 & 0.77 \\
		GOODSN & 2990 & 189.120080 & 62.156421 & 0.52 & 10.39 & 21.57 & 0.45 & 0.87 & 69.09 & 22.22 & 0.06 & 0.57 & -44.77 & 1.22 & 20.59 & 0.46 & 0.91 & 68.66 & 21.04 & 0.05 & 0.61 & -71.28 & 2.38 \\
		GOODSN & 3559 & 189.073042 & 62.162753 & 0.84 & 10.61 & 22.61 & 0.40 & 0.94 & -44.50 & 23.19 & 0.05 & 0.74 & 0.67 & 0.86 & 21.24 & 0.41 & 0.89 & -51.43 & 21.81 & 0.05 & 0.80 & 47.25 & 0.50 \\
		GOODSN & 5037 & 189.031491 & 62.176330 & 0.68 & 10.51 & 22.12 & 0.41 & 0.69 & 23.69 & 22.69 & 0.05 & 0.58 & -24.51 & 0.97 & 21.02 & 0.40 & 0.69 & 26.84 & 21.48 & 0.06 & 0.65 & -24.12 & 0.52 \\
		GOODSN & 5550 & 189.017861 & 62.180545 & 0.85 & 10.61 & 21.96 & 0.30 & 0.85 & 15.74 & 22.82 & 0.05 & 0.30 & -38.67 & 0.50 & 20.88 & 0.33 & 0.82 & 13.21 & 21.58 & 0.06 & 0.45 & -49.95 & 0.50 \\
		GOODSN & 8014 & 189.380631 & 62.197958 & 0.66 & 10.05 & 22.57 & 0.33 & 0.69 & 56.81 & 23.95 & 0.18 & 0.23 & 78.63 & 1.43 & 21.39 & 0.30 & 0.64 & 66.05 & 22.78 & 0.08 & 0.42 & 60.88 & 1.15 \\
		GOODSN & 8996 & 189.280274 & 62.203887 & 0.66 & 10.41 & 22.20 & 0.37 & 0.22 & 40.65 & 22.95 & 0.04 & 0.76 & 48.04 & 1.87 & 21.07 & 0.35 & 0.23 & 40.14 & 21.85 & 0.05 & 0.81 & 68.14 & 0.50 \\
		GOODSN & 9636 & 188.984243 & 62.208232 & 0.64 & 10.19 & 21.68 & 0.25 & 0.75 & 31.45 & 22.43 & 0.03 & 0.34 & 62.37 & 1.55 & 20.79 & 0.25 & 0.75 & 24.63 & 21.75 & 0.05 & 0.06 & 66.52 & 1.59 \\
		GOODSN & 10217 & 189.115841 & 62.211309 & 0.52 & 10.40 & 21.60 & 0.39 & 0.93 & 49.71 & 22.33 & 0.05 & 0.63 & -56.95 & 0.60 & 20.53 & 0.37 & 0.92 & 38.92 & 21.20 & 0.06 & 0.52 & -62.34 & 0.50 \\
		GOODSN & 15194 & 189.103741 & 62.244090 & 0.64 & 10.68 & 21.64 & 0.47 & 0.81 & -82.56 & 22.38 & 0.07 & 0.79 & -13.89 & 1.18 & 20.45 & 0.46 & 0.82 & -80.92 & 21.07 & 0.07 & 0.88 & -12.97 & 0.91 \\
		GOODSN & 16232 & 189.081045 & 62.251542 & 0.78 & 10.48 & 22.49 & 0.36 & 0.77 & -16.65 & 24.06 & 0.05 & 0.37 & -34.87 & 0.56 & 21.20 & 0.33 & 0.76 & -14.01 & 22.50 & 0.05 & 0.13 & -33.09 & 0.50 \\
		GOODSN & 16487 & 189.149117 & 62.253903 & 0.85 & 10.12 & 23.36 & 0.22 & 0.78 & -11.52 & 24.78 & 0.05 & 0.38 & -43.26 & 0.50 & 22.08 & 0.21 & 0.79 & -10.23 & 23.18 & 0.03 & 0.06 & -40.31 & 5.39 \\
		GOODSN & 18455 & 189.202558 & 62.264619 & 0.79 & 10.80 & 21.60 & 0.36 & 0.95 & -15.99 & 23.38 & 0.06 & 0.62 & 40.39 & 0.71 & 20.44 & 0.35 & 0.96 & -23.97 & 22.19 & 0.06 & 0.72 & 35.45 & 0.50 \\
		GOODSN & 19573 & 189.088501 & 62.271780 & 0.68 & 10.61 & 21.45 & 0.45 & 0.18 & -56.08 & 22.46 & 0.45 & 0.43 & 28.55 & 8.00 & 20.33 & 0.38 & 0.19 & -56.31 & 21.32 & 0.15 & 0.70 & 31.16 & 8.00 \\
		GOODSN & 22260 & 189.158314 & 62.291434 & 0.56 & 10.66 & 21.05 & 0.51 & 0.42 & -84.96 & 22.41 & 0.04 & 0.65 & -72.78 & 0.84 & 19.92 & 0.50 & 0.42 & -85.24 & 21.18 & 0.04 & 0.72 & -66.79 & 0.94 \\
		GOODSN & 23555 & 189.093608 & 62.307940 & 0.82 & 10.46 & 22.40 & 0.50 & 0.39 & -65.99 & 23.17 & 0.10 & 0.46 & -64.04 & 1.32 & 21.03 & 0.48 & 0.39 & -67.14 & 21.79 & 0.11 & 0.11 & -53.28 & 0.80 \\
		GOODSS & 1077 & 53.177959 & -27.917692 & 0.62 & 10.53 & 21.62 & 0.36 & 0.24 & 14.04 & 22.32 & 0.04 & 0.64 & 15.88 & 0.50 & 20.45 & 0.34 & 0.23 & 13.73 & 21.16 & 0.04 & 0.17 & 47.25 & 0.50 \\
		GOODSS & 1705 & 53.203255 & -27.904933 & 0.60 & 10.57 & 21.35 & 0.41 & 0.89 & -44.71 & 22.32 & 0.10 & 0.44 & 35.96 & 2.85 & 20.14 & 0.43 & 0.89 & -46.97 & 20.91 & 0.09 & 0.44 & 31.56 & 1.94 \\
		GOODSS & 5960 & 53.062416 & -27.857515 & 0.67 & 10.84 & 21.10 & 0.94 & 0.92 & -76.15 & 21.98 & 0.15 & 0.43 & -69.68 & 2.22 & 19.86 & 0.90 & 0.91 & -70.95 & 20.69 & 0.14 & 0.44 & -72.97 & 1.57 \\
		GOODSS & 9603 & 53.120834 & -27.823062 & 0.73 & 10.24 & 22.98 & 0.30 & 0.12 & -23.68 & 23.92 & 0.30 & 0.50 & -21.77 & 1.06 & 21.80 & 0.29 & 0.11 & -23.16 & 22.27 & 0.21 & 0.45 & -23.67 & 1.43 \\
		GOODSS & 11881 & 53.190754 & -27.803579 & 0.53 & 10.15 & 21.94 & 0.37 & 0.73 & -53.51 & 22.82 & 0.04 & 0.58 & -43.62 & 0.50 & 20.88 & 0.35 & 0.72 & -52.65 & 21.74 & 0.03 & 0.82 & -75.40 & 2.73 \\
		GOODSS & 12900 & 53.078293 & -27.795719 & 0.73 & 10.32 & 22.78 & 0.30 & 0.71 & 7.76 & 23.68 & 0.03 & 0.89 & 22.92 & 0.61 & 21.58 & 0.33 & 0.71 & 6.11 & 22.15 & 0.06 & 0.64 & 17.12 & 0.50 \\
		GOODSS & 13623 & 53.068878 & -27.790901 & 0.73 & 10.63 & 21.90 & 0.58 & 0.62 & 59.41 & 22.46 & 0.07 & 0.79 & 79.41 & 1.34 & 20.72 & 0.56 & 0.62 & 59.89 & 21.27 & 0.06 & 0.95 & 80.86 & 0.81 \\
		GOODSS & 19113 & 53.036057 & -27.750505 & 0.87 & 10.85 & 21.67 & 0.26 & 0.70 & -55.59 & 22.59 & 0.05 & 0.29 & -61.86 & 0.57 & 20.55 & 0.26 & 0.69 & -56.08 & 21.35 & 0.07 & 0.27 & 83.01 & 0.50 \\
		GOODSS & 19771 & 53.172154 & -27.743362 & 0.73 & 10.19 & 23.22 & 0.30 & 0.35 & 58.90 & 24.83 & 0.03 & 0.47 & -66.86 & 8.00 & 21.95 & 0.27 & 0.32 & 59.33 & 22.79 & 0.04 & 0.75 & 13.95 & 7.87 \\
		GOODSS & 20655 & 53.140907 & -27.736114 & 0.67 & 10.52 & 21.63 & 0.53 & 0.67 & -19.69 & 22.88 & 0.07 & 0.56 & 11.66 & 2.14 & 20.45 & 0.51 & 0.69 & -19.60 & 21.61 & 0.08 & 0.66 & 18.32 & 0.60 \\
		GOODSS & 20734 & 53.190987 & -27.735471 & 0.54 & 10.28 & 21.79 & 0.41 & 0.94 & 67.03 & 22.52 & 0.06 & 0.61 & 67.08 & 1.79 & 20.76 & 0.39 & 0.97 & 45.78 & 21.50 & 0.06 & 0.56 & 64.77 & 1.37 \\
		UDS & 6502 & 34.424271 & -5.240068 & 0.61 & 10.30 & 22.34 & 0.34 & 0.86 & 74.80 & 23.09 & 0.06 & 0.89 & 65.65 & 1.25 & 21.13 & 0.30 & 0.88 & 68.72 & 22.05 & 0.05 & 0.89 & -81.33 & 0.88 \\
		UDS & 8498 & 34.432415 & -5.230028 & 0.64 & 10.60 & 21.84 & 0.52 & 0.71 & -23.55 & 22.98 & 0.07 & 0.66 & 53.18 & 1.20 & 20.74 & 0.49 & 0.66 & -26.96 & 21.45 & 0.06 & 0.53 & 46.44 & 3.82 \\
		UDS & 8647 & 34.402882 & -5.228114 & 0.65 & 10.17 & 22.58 & 0.33 & 0.58 & -23.87 & 23.41 & 0.06 & 0.42 & -27.54 & 0.81 & 21.51 & 0.31 & 0.55 & -24.76 & 22.32 & 0.06 & 0.41 & -24.63 & 0.50 \\
		UDS & 9425 & 34.397026 & -5.225145 & 0.65 & 10.97 & 20.36 & 0.72 & 0.64 & -89.50 & 21.19 & 0.09 & 0.61 & -69.23 & 3.36 & 19.24 & 0.69 & 0.64 & -88.26 & 20.23 & 0.07 & 0.75 & -60.75 & 1.49 \\
		UDS & 11985 & 34.379379 & -5.210411 & 0.65 & 10.53 & 21.91 & 0.64 & 0.31 & 46.14 & 22.45 & 0.06 & 0.88 & 63.52 & 1.18 & 20.72 & 0.64 & 0.30 & 45.54 & 21.16 & 0.06 & 0.90 & 78.02 & 1.10 \\
		UDS & 12517 & 34.372631 & -5.207050 & 0.60 & 10.25 & 22.48 & 0.34 & 0.27 & -29.69 & 23.42 & 0.05 & 0.78 & -42.03 & 2.36 & 21.29 & 0.31 & 0.22 & -30.78 & 21.98 & 0.07 & 0.78 & 17.87 & 5.89 \\
		UDS & 13009 & 34.235287 & -5.205592 & 0.51 & 10.68 & 20.87 & 0.83 & 0.66 & 57.92 & 21.42 & 0.12 & 0.56 & 46.77 & 2.49 & 19.70 & 0.78 & 0.66 & 57.75 & 20.37 & 0.11 & 0.54 & 40.27 & 1.89 \\
		UDS & 19508 & 34.376335 & -5.169682 & 0.55 & 10.36 & 21.81 & 0.46 & 0.83 & -11.04 & 22.31 & 0.07 & 0.48 & -57.72 & 1.94 & 20.72 & 0.45 & 0.82 & -6.53 & 21.24 & 0.06 & 0.29 & -52.64 & 3.03 \\
		UDS & 20061 & 34.343971 & -5.166534 & 0.65 & 10.57 & 21.80 & 0.60 & 0.51 & 84.59 & 22.37 & 0.06 & 0.74 & 86.84 & 1.63 & 20.60 & 0.56 & 0.54 & 84.13 & 21.17 & 0.06 & 0.84 & -79.41 & 1.28 \\
		\hline
	\end{tabular}
	\label{tab:rs_galaxies}
\end{sidewaystable}

\end{document}